# On the Relative Strength of Pebbling and Resolution[*]

Jakob Nordström[†]

Computer Science and Artificial Intelligence Laboratory
Massachusetts Institute of Technology
Cambridge, MA 02139, USA
jakobn@mit.edu

November 7, 2018

## Abstract

The last decade has seen a revival of interest in pebble games in the context of proof complexity. Pebbling has proven to be a useful tool for studying resolution-based proof systems when comparing the strength of different subsystems, showing bounds on proof space, and establishing size-space trade-offs. The typical approach has been to encode the pebble game played on a graph as a CNF formula and then argue that proofs of this formula must inherit (various aspects of) the pebbling properties of the underlying graph. Unfortunately, the reductions used here are not tight. To simulate resolution proofs by pebblings, the full strength of nondeterministic black-white pebbling is needed, whereas resolution is only known to be able to simulate deterministic black pebbling. To obtain strong results, one therefore needs to find specific graph families which either have essentially the same properties for black and black-white pebbling (not at all true in general) or which admit simulations of black-white pebblings in resolution.

This paper contributes to both these approaches. First, we design a restricted form of black-white pebbling that can be simulated in resolution and show that there are graph families for which such restricted pebblings can be asymptotically better than black pebblings. This proves that, perhaps somewhat unexpectedly, resolution can strictly beat black-only pebbling, and in particular that the space lower bounds on pebbling formulas in [Ben-Sasson and Nordström 2008] are tight. Second, we present a versatile parametrized graph family with essentially the same properties for black and black-white pebbling, which gives sharp simultaneous trade-offs for black and black-white pebbling for various parameter settings. Both of our contributions have been instrumental in obtaining the time-space trade-off results for resolution-based proof systems in [Ben-Sasson and Nordström 2009].

## 1   Introduction

Pebbling is a tool for studying time-space relationships by means of a game played on directed acyclic graphs. This game models computations where the execution is independent of the input and can be performed by straight-line programs. Each such program is encoded as a graph, and a pebble on a vertex in the graph indicates that the corresponding value is currently kept in memory. The goal is to pebble the output vertex of the graph with minimal number of pebbles (amount of memory) and steps (amount of time).

Pebble games were originally devised for studying programming languages and compiler construction, but have later found a broad range of applications in computational complexity theory. The pebble game model seems to have appeared for the first time (implicitly) in [PH70], where it was used to study flowcharts and recursive schemata, and it was later employed to model register allocation [Set75], and analyze the relative power of time and space as Turing-machine resources [Coo74, HPV77]. Moreover, pebbling has been used to derive time-space trade-offs for algorithmic concepts such as linear recursion [Cha73, SS83], fast Fourier transform [SS77, Tom78], matrix multiplication [Tom78], and integer multiplication [SS79]. An excellent survey of pebbling up to ca 1980 is [Pip80], and some more recent developments are covered in the author's upcoming survey [Nor10a].

---

[*]This is the full-length version of the paper [Nor10b] to appear at the *25th IEEE Conference on Computational Complexity*.

[†]Research supported by the Royal Swedish Academy of Sciences, the Ericsson Research Foundation, the Sweden-America Foundation, the Foundation Olle Engkvist Byggmästare, the Sven and Dagmar Salén Foundation, and the Foundation Blanceflor Boncompagni-Ludovisi, née Bildt.

The *pebbling price* of a directed acyclic graph $G$ in the black pebble game captures the memory space, or number of registers, required to perform the deterministic computation described by $G$. We will mainly be interested in the more general *black-white pebble game* modelling nondeterministic computation, which was introduced in [CS76] and has been studied in [GT78, Kla85, LT80, LT82, Mey81, KS91, Wil88] and other papers.

**Definition 1.1 (Pebble game).** Let $G$ be a directed acyclic graph (DAG) with a unique sink vertex $z$. The *black-white pebble game* on $G$ is the following one-player game. At any time $t$, we have a configuration $\mathbb{P}_t = (B_t, W_t)$ of black pebbles $B_t$ and white pebbles $W_t$ on the vertices of $G$, at most one pebble per vertex. The rules of the game are as follows:

1. If all immediate predecessors of an empty vertex $v$ have pebbles on them, a black pebble may be placed on $v$. In particular, a black pebble can always be placed on a source vertex.

2. A black pebble may be removed from any vertex at any time.

3. A white pebble may be placed on any empty vertex at any time.

4. If all immediate predecessors of a white-pebbled vertex $v$ have pebbles on them, the white pebble on $v$ may be removed. In particular, a white pebble can always be removed from a source vertex.

A *(complete) black-white pebbling* of $G$, also called a *pebbling strategy for $G$*, is a sequence of pebble configurations $\mathcal{P} = \{\mathbb{P}_0, \ldots, \mathbb{P}_\tau\}$ such that $\mathbb{P}_0 = (\emptyset, \emptyset)$, $\mathbb{P}_\tau = (\{z\}, \emptyset)$, and for all $t \in [\tau]$, $\mathbb{P}_t$ follows from $\mathbb{P}_{t-1}$ by one of the rules above. The *time* of a pebbling $\mathcal{P} = \{\mathbb{P}_0, \ldots, \mathbb{P}_\tau\}$ is simply $\textit{time}(\mathcal{P}) = \tau$ and the *space* is $\textit{space}(\mathcal{P}) = \max_{0 \leq t \leq \tau}\{|B_t \cup W_t|\}$. The *black-white pebbling price* (also known as the *pebbling measure* or *pebbling number*) of $G$, denoted $\textit{BW-Peb}(G)$, is the minimum space of any complete pebbling of $G$.

A *black pebbling* is a pebbling using black pebbles only, i.e., having $W_t = \emptyset$ for all $t$. The *(black) pebbling price* of $G$, denoted $\textit{Peb}(G)$, is the minimum space of any complete black pebbling of $G$.

In the last decade, there has been renewed interest in pebbling in the context of proof complexity.[1] A (non-exhaustive) list of proof complexity papers using pebbling in one way or another is [AJPU07, BEGJ00, BIPS10, Ben09, BIW04, BN08, BN09a, BN09b, BW01, EGM04, ET01, ET03, HU07, Nor09, NH08b, SBK04]. The way pebbling results have been used in proof complexity has mainly been by studying so-called *pebbling contradictions* (also known as *pebbling formulas* or *pebbling tautologies*). These are CNF formulas encoding the pebble game played on a DAG $G$ by postulating the sources to be true and the sink to be false, and specifying that truth propagates through the graph according to the pebbling rules. The idea to use such formulas seems to have appeared for the first time in [Koz77], and they were also studied in [RM99, BEGJ00] before being explicitly defined in [BW01].

**Definition 1.2 (Pebbling contradiction).** Suppose that $G$ is a DAG with sources $S$ and a unique sink $z$. Identify every vertex $v \in V(G)$ with a propositional logic variable $v$. The *pebbling contradiction* over $G$, denoted $Peb_G$, is the conjunction of the following clauses:

---

[1]We remark that the pebble game studied in this paper should not be confused with the (very different) *existential pebble games* that have also been used in proof complexity, for instance, in the papers [Ats04, AD08, AKV04, BG03, GT05].

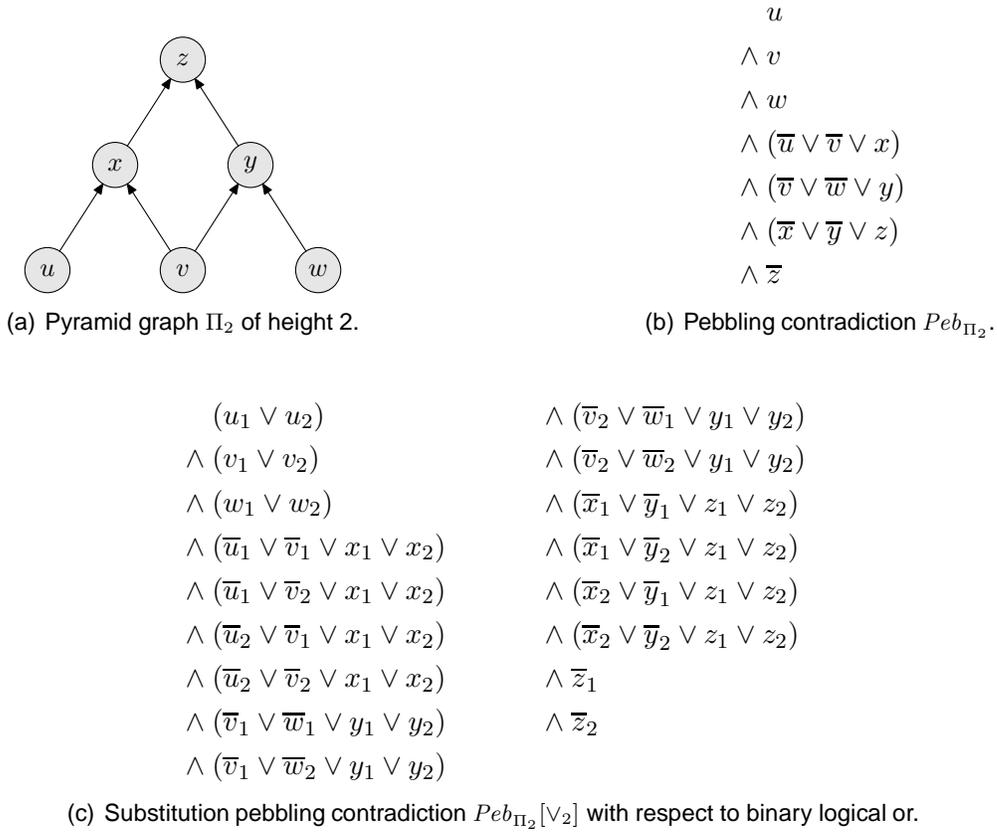

(a) Pyramid graph $\Pi_2$ of height 2.

(b) Pebbling contradiction $Peb_{\Pi_2}$.

$$
\begin{aligned}
&(u_1 \vee u_2) && \wedge\, (\overline{v}_2 \vee \overline{w}_1 \vee y_1 \vee y_2)\\
\wedge\,&(v_1 \vee v_2) && \wedge\, (\overline{v}_2 \vee \overline{w}_2 \vee y_1 \vee y_2)\\
\wedge\,&(w_1 \vee w_2) && \wedge\, (\overline{x}_1 \vee \overline{y}_1 \vee z_1 \vee z_2)\\
\wedge\,&(\overline{u}_1 \vee \overline{v}_1 \vee x_1 \vee x_2) && \wedge\, (\overline{x}_1 \vee \overline{y}_2 \vee z_1 \vee z_2)\\
\wedge\,&(\overline{u}_1 \vee \overline{v}_2 \vee x_1 \vee x_2) && \wedge\, (\overline{x}_2 \vee \overline{y}_1 \vee z_1 \vee z_2)\\
\wedge\,&(\overline{u}_2 \vee \overline{v}_1 \vee x_1 \vee x_2) && \wedge\, (\overline{x}_2 \vee \overline{y}_2 \vee z_1 \vee z_2)\\
\wedge\,&(\overline{u}_2 \vee \overline{v}_2 \vee x_1 \vee x_2) && \wedge\, \overline{z}_1\\
\wedge\,&(\overline{v}_1 \vee \overline{w}_1 \vee y_1 \vee y_2) && \wedge\, \overline{z}_2\\
\wedge\,&(\overline{v}_1 \vee \overline{w}_2 \vee y_1 \vee y_2)
\end{aligned}
$$

(c) Substitution pebbling contradiction $Peb_{\Pi_2}[\vee_2]$ with respect to binary logical or.

**Figure 1:** Example of pebbling contradiction with substitution for the pyramid graph $\Pi_2$.

- for all $s \in S$, a unit clause $s$ (*source axioms*),

- For all non-sources $v$ with immediate predecessors $pred(v)$, the clause $\bigvee_{u \in pred(v)} \overline{u} \vee v$ (*pebbling axioms*),

- for the sink $z$, the unit clause $\overline{z}$ (*target* or *sink axiom*).

For any nonconstant Boolean function $f_d : \{0,1\}^d \mapsto \{0,1\}$, the *substitution pebbling contradiction with respect to* $f_d$ is the CNF formula $Peb_G[f_d]$ obtained by substituting $f_d(x_1, \ldots, x_d)$ for every variable $x$ and expanding the result to conjunctive normal form in the canonical way.

If the graph $G$ has $n$ vertices and maximal indegree $\ell$, $Peb_G[f_d]$ is easily verified to be an unsatisfiable formula over $dn$ variables with less than $2^{d(\ell+1)} \cdot n$ clauses of size at most $d(\ell+1)$. An example illustrating Definition 1.2 is given in Figure 1.

Given any black-only pebbling $\mathcal{P}$ of $G$, it is straightforward to simulate this pebbling in resolution to refute the corresponding pebbling contradiction $Peb_G[f_d]$ in length $\mathrm{O}\big(\textit{time}(\mathcal{P})\big)$ and space $\mathrm{O}\big(\textit{space}(\mathcal{P})\big)$. This was perhaps first noted in [BIW04] for the simple $Peb_G$ formulas, but the simulation extends readily to any formula $Peb_G[f_d]$, with the constants hidden in the asymptotic notation depending on $f_d$ and the maximal indegree of $G$. In the other direction, it was recently shown in [BN09b] (strengthening results in [BN08]) that if $f_d$ has the right properties—for instance, if it is the exclusive or function or the threshold function evaluating to true if $k$ out of $d$ variables are true for $1 < k < d$—then any resolution refutation $\pi$

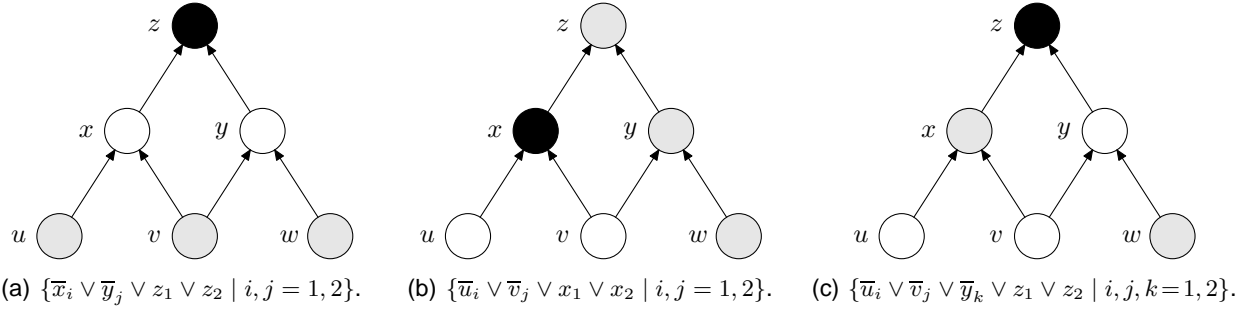

(a) $\{\overline{x}_i \vee \overline{y}_j \vee z_1 \vee z_2 \mid i,j=1,2\}.$

(b) $\{\overline{u}_i \vee \overline{v}_j \vee x_1 \vee x_2 \mid i,j=1,2\}.$

(c) $\{\overline{u}_i \vee \overline{v}_j \vee \overline{y}_k \vee z_1 \vee z_2 \mid i,j,k=1,2\}.$

**Figure 2:** Black and white pebbles and (intuitively) corresponding sets of clauses.

of $Peb_G[f_d]$ can be translated into a black-white pebbling of $G$ with time and space upper-bounded by the length and space of $\pi$, respectively (adjusted for small multiplicative constants depending on the maximal indegree of $G$).

There is an obvious gap in these reductions between pebbling and resolution. To interpret a resolution refutation of a pebbling contradiction in terms of a pebbling of the underlying graph, the full power of black-white pebbling is needed to make the reduction work. If we want to translate pebblings of graphs into refutations of the corresponding pebbling contradictions, however, we only know how to do this for the weaker black pebble game.

To see why resolution has a hard time simulating black-white pebblings, let us start by discussing a black-only pebbling $\mathcal{P}$. We can easily mimic such a pebbling in a resolution refutation of $Peb_G[f_d]$ by deriving that $f_d(v_1, \ldots, v_d)$ is true whenever the corresponding vertex $v$ in $G$ is black-pebbled. We end up deriving that $f_d(z_1, \ldots, z_d)$ is true for the sink $z$, at which point we can download the sink axioms and derive a contradiction. The intuition behind this translation is that a black pebble on $v$ means that we know $v$, which in resolution translates into truth of $v$. In the pebble game, having a white pebble on $v$ instead means that we need to assume $v$. By duality, we let this correspond to falsity of $v$ in resolution. Focusing on the pyramid $\Pi_2$ and pebbling contradiction $Peb_{\Pi_2}[\vee_2]$ in Figure 1, our intuitive understanding then becomes that white pebbles on $x$ and $y$ and a black pebble on $z$ should correspond to the set of clauses

$$\{\overline{x}_i \vee \overline{y}_j \vee z_1 \vee z_2 \mid i,j=1,2\} \tag{1.1}$$

which indeed encode that assuming $x_1 \vee x_2$ and $y_1 \vee y_2$, we can deduce $z_1 \vee z_2$. See Figure 2(a) for an illustration of this.

If we now place white pebbles on $u$ and $v$, this allows us to remove the white pebble from $x$. Rephrasing this in terms of resolution, we can say that $x$ follows if we assume $u$ and $v$, which is encoded as the set of clauses

$$\{\overline{u}_i \vee \overline{v}_j \vee x_1 \vee x_2 \mid i,j=1,2\} \tag{1.2}$$

(see Figure 2(b)), and indeed, from the clauses in (1.1) and (1.2) we can derive in resolution that $z$ is black-pebbled and $u$, $v$ and $y$ are white pebbled, i.e., the set of clauses

$$\{\overline{u}_i \vee \overline{v}_j \vee \overline{y}_k \vee z_1 \vee z_2 \mid i,j,k=1,2\} \tag{1.3}$$

(see Figure 2(c)). This toy example indicates one of the problems one runs into when one tries to simulate black-white pebbling in resolution: as the number of white pebbles grows, there is an exponential blow-up in the number of clauses. The clause set in (1.3) is twice the size of those in (1.1) and (1.2), although it corresponds to only one more white pebble. This suggests that as a pebbling starts to make heavy use of white pebbles, a resolution refutation will not be able to mimic such a pebbling in a length- and space-preserving manner.

This leads to the thought that perhaps black pebbling provides not only upper but also lower bounds on resolution refutations of pebbling contradictions. This would be consistent with what has been known so far. For all pebbling contradictions with proven space lower bounds, the underlying graphs have asymptotically the same black and black-white pebbling price, and hence all known lower bounds can be expressed in terms of black pebbling. There have been no examples of pebbling contradictions where resolution can do strictly better than black pebbling and tightly match smaller bounds on space in terms of black-white pebbling.

## 1.1   Our Results

Our first set of results is that resolution can in fact be strictly better than black-only pebbling, both for time-space trade-offs and with respect to space in absolute terms. We prove this by designing a limited version of black-white pebbling, where we explicitly restrict the amount of nondeterminism, i.e., white pebbles, a pebbling strategy can use. Such restricted pebbling use "few white pebbles per black pebble" (in a sense that will be made formal below), and can therefore be simulated in a time- and space-preserving manner by resolution, avoiding the exponential blow-up just discussed. We then show that for all known separation results in the pebbling literature where black-white pebbling does asymptotically better than black-only pebbling, there are graphs exhibiting these separations for which optimal black-white pebblings can be carried out in our limited version of the game. This means that resolution refutations of pebbling contradictions over such DAGs can do strictly asymptotically better than what is suggested by black-only pebbling, matching the lower bounds in terms of (general) black-white pebbling.

More precisely, we obtain such results for three families of graphs.[2] The first family are the *bit reversal graphs* studied by Lengauer and Tarjan [LT82], for which black-white pebbling has quadratically better trade-offs than black pebbling. (We refer to Section 3 for all formal notation and definitions used below.)

**Lemma 1.3 ([LT82]).** *There are DAGs $\{G_n\}_{n=1}^{\infty}$ of size $\Theta(n)$ with black pebbling price $\mathsf{Peb}(G_n) = 3$ such that any optimal black pebbling $\mathcal{P}_n$ of $G_n$ exhibits a trade-off $\mathsf{time}(\mathcal{P}_n) = \Theta\big(n^2/\mathsf{space}(\mathcal{P}_n) + n\big)$ but optimal black-white pebblings $\mathcal{P}_n$ of $G_n$ achieve a trade-off $\mathsf{time}(\mathcal{P}_n) = \Theta\big((n/\mathsf{space}(\mathcal{P}_n))^2 + n\big)$.*

**Theorem 1.4.** *Fix any non-constant Boolean function $f$ and let $Peb_{G_n}[f]$ be pebbling contradictions over the graphs in Lemma 1.3. Then for any monotonically nondecreasing function $s(n) = \mathrm{O}(\sqrt{n})$ there are resolution refutations $\pi_n$ of $Peb_{G_n}[f]$ in total space $\mathrm{O}(s(n))$ and length $\mathrm{O}\big((n/s(n))^2\big)$, beating the lower bound $\Omega\big(n^2/s(n)\big)$ for black-only pebblings of $G_n$.*

Focusing next on absolute bounds on space rather than time-space trade-offs, the best known separation between black and black-white pebbling for polynomial-size graphs is the one shown by Wilber [Wil88].

**Lemma 1.5 ([Wil88]).** *There are DAGs $\{G(s)\}_{s=1}^{\infty}$ of size polynomial in $s$ with black-white pebbling price $\mathsf{BW\text{-}Peb}(G(s)) = \mathrm{O}(s)$ and black pebbling price $\mathsf{Peb}(G(s)) = \Omega(s \log s / \log \log s)$.*

For pebbling formulas over these graphs we do *not* know how to beat the black pebbling space bound— we return to this somewhat intriguing problem in Section 7—but using instead graphs in [KS91] exhibiting the same pebbling properties, we can obtain the desired result.

**Theorem 1.6.** *Fix any non-constant Boolean function $f$ and let $Peb_{G(s)}[f]$ be pebbling contradictions over the graphs $G(s)$ in [KS91] with pebbling properties as in Lemma 1.5. Then there are resolution refutations $\pi_n$ of $Peb_{G(s)}[f]$ in total space $\mathrm{O}(s)$, beating the lower bound $\Omega(s \log s / \log \log s)$ for black-only pebbling.*

If we remove all restriction on graph size, there is a quadratic separation of black and black-white pebbling established by Kalyanasundaram and Schnitger [KS91].

---

[2]All graphs discussed in this paper are explicitly constructible and have bounded vertex indegree. Also, unless otherwise stated they have a single, unique sink. We do not repeat this in the formal statements here in order not to clutter the text unnecessarily.

**Lemma 1.7 ([KS91]).** *There are DAGs $\{G(s)\}_{s=1}^{\infty}$ of size $\exp(\Theta(s \log s))$ such that $BW\text{-}Peb(G(s)) \leq 3s + 1$ but $Peb(G(s)) \geq s^2$.*

For pebbling formulas over these graphs, resolution again matches the black-white pebbling bounds.

**Theorem 1.8.** *Fix any non-constant Boolean function $f$ and let $Peb_{G(s)}[f]$ be pebbling contradictions over the graphs $G(s)$ in Lemma 1.7. Then there are resolution refutations $\pi_n$ of $Peb_{G(s)}[f]$ in total space $O(s)$, beating the lower bound $\Omega(s^2)$ for black-only pebbling.*

In particular, Theorems 1.6 and 1.8 show that the lower bound on proof space for pebbling contradictions in terms of black-white pebbling price in [BN08] is tight (up to constant factors).

Turning to our second set of results, we first note that in spite of the theorems above, for general pebbling formulas we still do not know of any way of simulating black-white pebbling in resolution. Instead, we are limited to deriving upper bounds from black-only pebblings while lower bounds have to be obtained in terms of black-white pebblings. At first sight, this might not look too bad since the space gap between the two can be at most quadratic, as shown by Meyer auf der Heide [Mey81]. However, the translation given in [Mey81] of a black-white pebbling in space $s$ to a black pebbling in space $O(s^2)$ incurs an exponential blow-up in pebbling time, destroying all hope of obtaining nontrivial time-space trade-off results for resolution in this way. Hence, to get meaningful trade-offs for pebbling formulas we need graph families with strong *dual* trade-offs for black and black-white pebbling simultaneously. In this paper, we present such a family of graphs, building on and strengthening previous work by Carlson and Savage [CS80, CS82].

**Theorem 1.9.** *There is an explicitly constructible two-parameter graph family $\Gamma(c, r)$, for $c, r \in \mathbb{N}^+$, having unique sink, vertex indegree 2, and size $\Theta(cr^3 + c^3r^2)$, and satisfying the following properties:*

1. *$\Gamma(c, r)$ has black-white pebbling price $BW\text{-}Peb(\Gamma(c, r)) = r + O(1)$ and black pebbling price $Peb(\Gamma(c, r)) = 2r + O(1)$.*

2. *There is a black-only pebbling of $\Gamma(c, r)$ in time linear in the graph size and in space $O(c + r)$.*

3. *Suppose that $\mathcal{P}$ is a black-white pebbling of $\Gamma(c, r)$ with $space(\mathcal{P}) \leq r + s$ for $0 < s \leq c/8$. Then $time(\mathcal{P}) \geq \left(\frac{c - 2s}{4s + 4}\right)^r \cdot r!$.*

The graph family in Theorem 1.9 turns out to be surprisingly versatile. For instance, we can use it to prove among other things the rather striking statement that for any *arbitrarily slowly growing* non-constant function, there are explicit graphs of such (arbitrarily small) pebbling space complexity that nevertheless exhibit *superpolynomial* time-space trade-offs for black and black-white pebbling simultaneously.

**Theorem 1.10.** *Let $g(n)$ be any arbitrarily slowly growing[3] monotone function $\omega(1) = g(n) = O(n^{1/7})$, and let $\epsilon > 0$ be an arbitrarily small positive constant. Then there is a family of explicitly constructible single-sink DAGs $\{G_n\}_{n=1}^{\infty}$ of size $\Theta(n)$ such that the following holds:*

1. *The graph $G_n$ has black-white pebbling price $BW\text{-}Peb(G) = g(n) + O(1)$ and black pebbling price $Peb(G) = 2 \cdot g(n) + O(1)$.*

2. *There is a complete black pebbling $\mathcal{P}$ of $G_n$ with $time(\mathcal{P}) = O(n)$ and $space(\mathcal{P}) = O\left(\sqrt[3]{n/g^2(n)}\right)$*

3. *Any complete black-white pebbling of $G_n$ in space at most $\left(n/g^2(n)\right)^{1/3-\epsilon}$ requires pebbling time superpolynomial in $n$.*

More examples of interesting trade-offs that can be obtained from the graphs in Theorem 1.9 are given in Section 6.

---

[3] Note that we also assume $g(n) = O(n^{1/7})$, i.e., that $g(n)$ does not grow to fast. This is just a simplifying technical assumption. If we allow the minimal space to grow as fast as $n^\epsilon$ for some $\epsilon > 0$, then it is easy to use our graph family with other parameter settings to obtain even stronger results. Hence, the interesting aspect here is that $g(n)$ is allowed to grow arbitrarily slowly.

## 1.2   Organization of This Paper

In Section 2 we outline the main ideas behind our results, and Section 3 provides all the necessary pre-liminaries for the formal proofs of these results given in the rest of the paper. Section 4 proves our claims about the limited type of black-white pebblings that can be simulated by resolution, and in Section 5 we show that there are such limited pebblings for some interesting graph families. In Section 6, we discuss the graphs exhibiting our new pebbling trade-off results, and show how different parameter settings yield strong dual time-space trade-offs with upper bounds for black pebbling and matching lower bounds for black-white pebbling. We conclude in Section 7 by discussing some remaining open problems.

# 2   Outline of Constructions and Proofs

We will need to set up a fair amount of technical machinery before we can give the full, formal proofs of our results. In order not to obscure unnecessarily what are in essence reasonably straightforward arguments, in this section we try to give an overview of the main ideas, postponing the technicalities for later.

## 2.1   Limited Black-White Pebblings That Can Be Simulated by Resolution

Let us start by discussing the tools used to establish Theorems 1.4, 1.6, and 1.8. The idea is to design a version of the black-white pebble game that is tailor-made for resolution. This game is essentially just a for-malization of the naive resolution simulation sketched in Section 1, but before stating the formal definitions, let us try to provide some intuition why the rules of this new game look the way they do.

First, if we want a game that can be mimicked by resolution, then placements of isolated white vertices do not make much sense. What a resolution derivation can do is to download axiom clauses, and intuitively this corresponds to placing a black pebble on a vertex together with white pebbles on its immediate pre-decessors, if it has any. Therefore, we adopt such aggregate moves as the only admissible way of placing new pebbles. For instance, looking at the graph $\Pi_2$ and pebbling contradiction $Peb_{\Pi_2}[\vee_2]$ in Figure 1 again, placing a black pebble on $z$ and white pebbles on $x$ and $y$ corresponds to downloading the axiom clauses in (1.1).

Second, note that if we have a black pebble on $z$ with white pebbles on $x$ and $y$ corresponding to the clauses in (1.1) and a black pebble on $x$ with white pebbles on $u$ and $v$ corresponding to the clauses in (1.2), we can derive the clauses in (1.3) corresponding to $z$ black-pebbled and $u$, $v$ and $y$ white-pebbled but no pebble on $x$. This suggests that a natural rule for white pebble removal is that a white pebble can be removed from a vertex if a black pebble is placed *on that same vertex* (and not on its immediate predecessors).

Third, if we then just erase all clauses in (1.3), this corresponds to all pebbles disappearing. On the face of it, this is very much unlike the rule for white pebble removal in the standard pebble game, where it is absolutely crucial that a white pebble can only be removed when its predecessors are pebbled. However, the important point here is that not only do the white pebbles disappear—the black pebble that has been placed on $z$ with the help of these white pebbles disappears as well. What this means is that we cannot treat black and white pebbles in isolation, but we have to keep track of for each black pebble which white pebbles it depends on, and make sure that the black pebble also is erased if any of the white pebbles supporting it is erased. The way we do this is to label each black pebble $v$ with its supporting white pebbles $W$, and define the pebble game in terms of moves of such labelled *pebble subconfigurations* $v\langle W \rangle$.

**Definition 2.1 (Pebble subconfiguration).** For $v$ a vertex and $W$ a set of vertices, we say that $v\langle W \rangle$ is a *pebble subconfiguration* with a black pebble on $v$ supported by white pebbles on $W$. The black pebble on $v$ is said to be *dependent* on the white pebbles in its *support* $W$. We refer to $v\langle \emptyset \rangle$ as an *independent black pebble*.

Our next definition now formalizes the informal description of our new pebble game. We remark that this definition is quite similar to the pebble game defined in [Nor09], and that we have borrowed freely from notation and terminology there.

**Definition 2.2 (Labelled pebbling).** For $G$ any DAG with unique sink $z$, a (complete) *labelled pebbling* of $G$ is a sequence $\mathcal{L} = \{\mathbb{L}_0, \ldots, \mathbb{L}_\tau\}$ of labelled pebble configurations such that $\mathbb{L}_0 = \emptyset$, $\mathbb{L}_\tau = \{z\langle\emptyset\rangle\}$, and for all $t \in [\tau]$ it holds that $\mathbb{L}_t$ can be obtained from $\mathbb{L}_{t-1}$ by one of the following rules:

***Introduction*** $\mathbb{L}_t = \mathbb{L}_{t-1} \cup \{v\langle pred(v)\rangle\}$, where $pred(v)$ is the set of immediate predecessors of $v$.

***Erasure*** $\mathbb{L}_t = \mathbb{L}_{t-1} \setminus \{v\langle V\rangle\}$ for $v\langle V\rangle \in \mathbb{L}_{t-1}$.

***Merger*** $\mathbb{L}_t = \mathbb{L}_{t-1} \cup \{v\langle (V \cup W) \setminus \{w\}\rangle\}$ for $v\langle V\rangle, w\langle W\rangle \in \mathbb{L}_{t-1}$ with $w \in V$. We denote this subconfiguration $\mathsf{merge}(v\langle V\rangle, w\langle W\rangle)$, and refer to it as a *merger on* $w$.

Let $Bl(\mathbb{L}_t) = \bigcup \{v \mid v\langle W\rangle \in \mathbb{L}_t\}$ denote the set of all black-pebbled vertices in $\mathbb{L}_t$ and $Wh(\mathbb{L}_t) = \bigcup \{W \mid v\langle W\rangle \in \mathbb{L}_t\}$ the set of all white-pebbled vertices. Then the space of an labelled pebbling $\mathcal{L} = \{\mathbb{L}_0, \ldots, \mathbb{L}_\tau\}$ is $\max_{\mathbb{L} \in \mathcal{L}} \{|Bl(\mathbb{L}) \cup Wh(\mathbb{L})|\}$ and the time of $\mathcal{L}$ is $\textit{time}(\mathcal{L}) = \tau$.

Figures 2(a) and 2(b) are both examples of subconfigurations resulting from introduction moves, and if we merge the two we get the subconfiguration in Figure 2(c).

The game in Definition 2.2 might look quite different from the standard black-white pebble game, but it is not hard to show that labelled pebblings are essentially just a restricted form of black-white pebblings. (The proof of this is deferred to Section 4.)

**Lemma 2.3.** *If $G$ is a single-sink DAG and $\mathcal{L}$ is a complete labelled pebbling of $G$, then there is a complete black-white pebbling $\mathcal{P}_\mathcal{L}$ of $G$ with $\textit{time}(\mathcal{P}_\mathcal{L}) \leq \frac{4}{3}\textit{time}(\mathcal{L})$ and $\textit{space}(\mathcal{P}_\mathcal{L}) \leq \textit{space}(\mathcal{L})$.*

However, the definition of space of labelled pebblings does not seem quite right from the point of view of resolution. Not only does the space measure fail to capture the exponential blow-up in the number of white pebbles discussed above. We also have the problem that if one white pebble is used to support many different black pebbles, then in a resolution refutation simulating such a pebbling we have to pay multiple times for this single white pebble, once for every black pebble supported by it. To get something that can be simulated by resolution, we therefore need to restrict the labelled pebble game even further.

**Definition 2.4 (Bounded labelled pebblings).** An $(m, S)$-*bounded labelled pebbling* is a labelled pebbling $\mathcal{L} = \{\mathbb{L}_0, \ldots, \mathbb{L}_\tau\}$ such that every $\mathbb{L}_t$ contains at most $m$ pebble subconfigurations $v\langle W\rangle$ and every $v\langle W\rangle$ has white support size $|W| \leq S$.

Observe that boundedness automatically implies low space complexity, since an $(m, S)$-bounded pebbling $\mathcal{L}$ clearly satisfies $\textit{space}(\mathcal{L}) \leq m(S+1)$. And using the concept of bounded labelled pebblings, we can show that if there is such a pebbling of a graph $G$, then this pebbling can be used as a template for a resolution refutation of any pebbling contradiction $Peb_G[f]$. (We again refer to Section 4 for the proof.)

**Lemma 2.5.** *Suppose that $\mathcal{L}$ is any complete $(m, S)$-bounded pebbling of a graph $G$ and that $f$ is any nonconstant Boolean function of arity $d$. Then there is a resolution refutation $\pi_\mathcal{L}$ of the formula $Peb_G[f]$ in simultaneous length $L(\pi_\mathcal{L}) = \textit{time}(\mathcal{L}) \cdot \exp(\mathrm{O}(dS))$ and total space $TotSp(\pi_\mathcal{L}) = m \cdot \exp(\mathrm{O}(dS))$. In particular, fixing $f$ it holds that resolution can simulate $(m, \mathrm{O}(1))$-bounded pebblings in a time- and space-preserving manner.*

The whole problem thus boils down to the question whether there are graphs with (a) asymptotically different properties for black and black-white pebbling for which (b) optimal black-white pebblings can be carried out in the bounded labelled pebbling framework. The answer to this question turns out to be yes, and the space upper bounds for the pebbling contradictions in Theorems 1.4, 1.6, and 1.8 are all proven by exhibiting bounded labelled pebblings for the corresponding graphs. The details concerning how these graphs are constructed, as well as how they are pebbled, are somewhat intricate, however, and are therefore presented separately in Section 5.

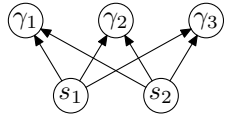

**Figure 3:** Base case for Carlson-Savage graph with $3$ spines and sinks.

## 2.2 A Graph Family with Tight Trade-offs for Black and Black-White Pebbling

Let us next outline the proof of our graph pebbling trade-off results in Theorem 1.9. We remark that in what follows, we will discuss a slightly different setting where graphs may have multiple sinks, not just one, and where we only require that a pebbling *visits* every sink once, touching it with a black or white pebble, instead of leaving a black pebble on the sink until the end of the pebbling. It is straightforward to translate results for such pebblings back to the setting in Theorem 1.9. (See Section 3 for the technical details.)

Our graph family is built on a construction by Carlson and Savage [CS80, CS82]. Carlson and Savage only prove their trade-off for black pebbling, however, and the extension of their results to black-white pebbling requires changing the construction and doing a nontrivial amount of extra work (as is usually the case when one wants to lift a black pebbling result to black-white pebbling). The formal definition of the family of graphs, which we will refer to as *Carlson-Savage graphs*, is probably easier to parse if the reader first studies the illustrations in Figures 3 and 4.

**Definition 2.6 (Carlson-Savage graphs).** The two-parameter graph family $\Gamma(c, r)$, for $c, r \in \mathbb{N}^+$, is defined by induction over $r$. The base case $\Gamma(c, 1)$ is a DAG consisting of two sources $s_1, s_2$ and $c$ sinks $\gamma_1, \ldots, \gamma_c$ with directed edges $(s_i, \gamma_j)$, for $i = 1, 2$ and $j = 1, \ldots, c$, i.e., edges from both sources to all sinks. The graph $\Gamma(c, r + 1)$ has $c$ sinks and is built from the following components:

- $c$ disjoint copies $\Pi_{2r}^{(1)}, \ldots, \Pi_{2r}^{(c)}$ of a pyramid graph[4] of height $2r$ with sinks $z_1, \ldots, z_c$.

- one copy of $\Gamma(c, r)$, for which we denote the sinks by $\gamma_1, \ldots, \gamma_c$.

- $c$ disjoint and identical *spines*, where each spine is composed of $cr$ *sections*, and every section contains $2c$ vertices. We let the vertices in the $i$th section of a spine be denoted $v[i]_1, \ldots, v[i]_{2c}$.

The edges in $\Gamma(c, r + 1)$ are as follows:

- All "internal edges" in $\Pi_{2r}^{(1)}, \ldots, \Pi_{2r}^{(c)}$ and $\Gamma(c, r)$ are present also in $\Gamma(c, r + 1)$.

- For each spine, there are edges $\left(v[i]_j, v[i]_{j+1}\right)$ for all $j = 1, \ldots, 2c - 1$ within each section $i$ and edges $\left(v[i]_{2c}, v[i + 1]_1\right)$ from the end of a section to the beginning of next for $i = 1, \ldots, cr - 1$, i.e., for all sections but the final one, where $v[cr]_{2c}$ is a sink.

- For each section $i$ in each spine, there are edges $\left(z_j, v[i]_j\right)$ from the $j$th pyramid sink to the $j$th vertex in the section for $j = 1, \ldots, c$, as well as edges $\left(\gamma_j, v[i]_{c+j}\right)$ from the $j$th sink in $\Gamma(c, r)$ to the $(c + j)$th vertex in the section for $j = 1, \ldots, c$.

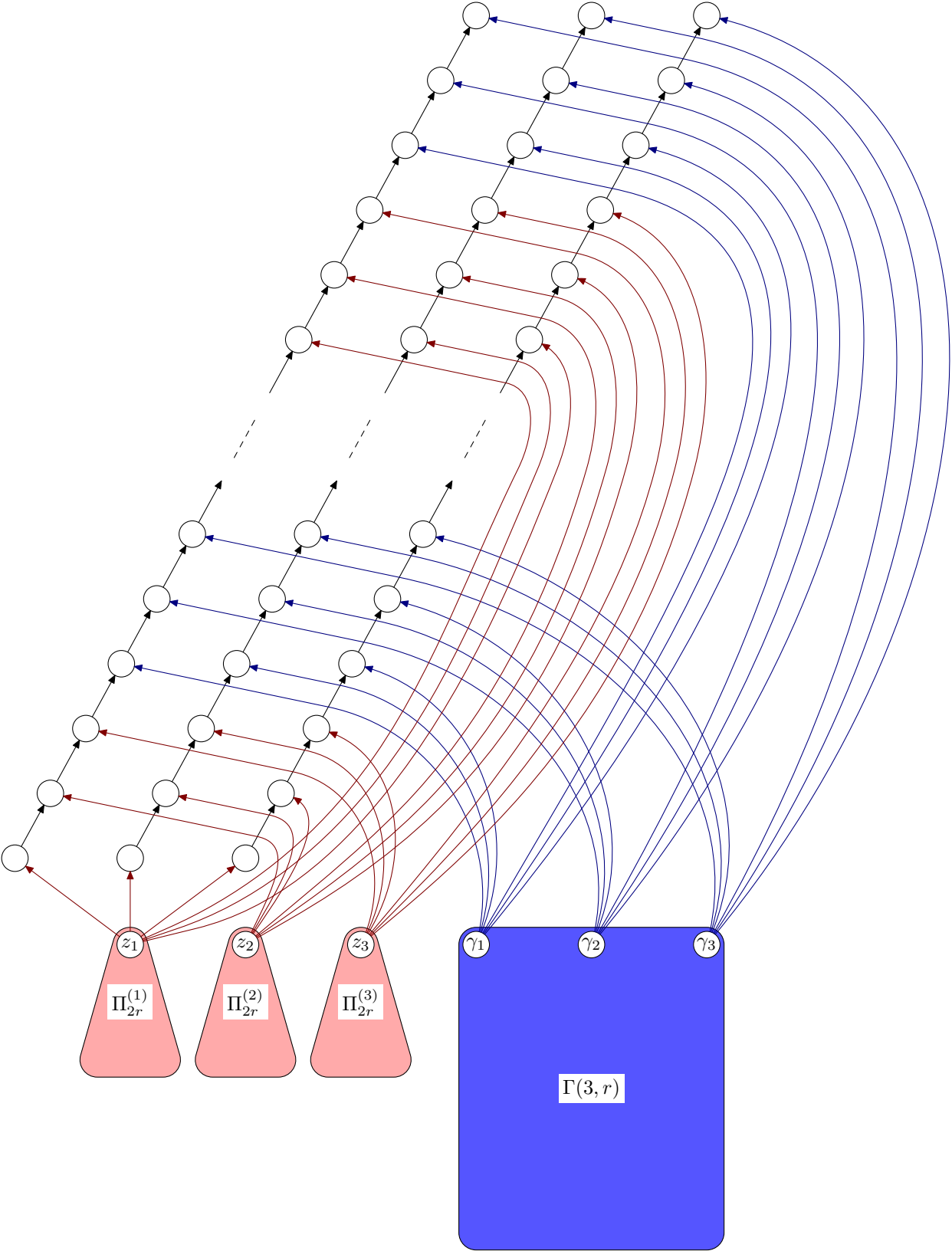

**Figure 4:** Inductive definition of Carlson-Savage graph $\Gamma(3, r+1)$ with $3$ spines and sinks.

Let us focus on the trade-off lower bound in part 3 of Theorem 1.9, which is the hard part to prove, and let us start by trying to provide some intuition why this bound should hold. For simplicity, consider first black-only pebblings. Assume inductively that part 3 of Theorem 1.9 has been proven for $\Gamma(c, r-1)$ and consider $\Gamma(c, r)$. Any pebbling strategy for this DAG will have to pebble through all sections in all spines. Consider the first section anywhere, let us say on spine $j$, that has been completely pebbled, i.e., there have been pebbles placed on and removed from all vertices in the section. Let us say that this happens at time $\tau_1$. But this means that $\Gamma(c, r-1)$ and all pyramids $\Pi_{2(r-1)}^{(1)}, \ldots, \Pi_{2(r-1)}^{(c)}$ must have been completely pebbled during this part of the pebbling as well. Fix any pyramid and consider some point in time $\sigma_1 < \tau_1$ when there are at least $r+1$ pebbles on its vertices, which must happen because of known pebbling lower bounds for pyramids [Coo74, Kla85]. At this point, the rest of the graph must contain very few pebbles (think of $s$ here as being very small). In particular, there are very few pebbles on the subgraph $\Gamma(c, r-1)$ at time $\sigma_1$, so for all practical purposes we can think of $\Gamma(c, r-1)$ as being essentially empty of pebbles.

Consider now the next section in the spine $j$ that is completed, say, at time $\tau_2 > \tau_1$. Again, we can argue that some pyramid is completely pebbled in the time interval $[\tau_1, \tau_2]$, and thus has $r+1$ pebbles on it at some time $\sigma_2 > \tau_1 > \sigma_1$. This means that $\Gamma(c, r-1)$ is essentially empty of pebbles at time $\sigma_2$ as well. But note that all sinks in the subgraph $\Gamma(c, r-1)$ must have been pebbled in the time interval $[\sigma_1, \sigma_2]$, and since we know that $\Gamma(c, r-1)$ is (almost) empty at times $\sigma_1$ and $\sigma_2$, this allows us to apply the induction hypothesis. Since $\mathcal{P}$ has to pebble through a lot of sections in different spines, we will be able to repeat the above argument many times and apply the induction hypothesis on $\Gamma(c, r-1)$ each time. Adding up all the lower bounds obtained in this way, the induction step goes through.

This is the spirit of the proof of the black-only pebbling trade-off in [CS82]. When we instead want to deal with black-white pebblings, things get much more complicated. Black pebblings must by necessity pebble through a graph in a bottom-up fashion, and it is therefore straightforward to measure "how far" a black pebbling has progressed. A black-white pebbling, however, can place and remove pebbles anywhere in the DAG at any time. Therefore, it is more difficult to control the progress of a black-white pebbling, and one has to use different ideas and work harder in the proof.

We establish part 3 of Theorem 1.9 by proving a slightly stronger lemma, dealing with *conditional* pebblings that start with some pebbles already present on the graph, and can also leave some pebbles on the graph at the end of the pebbling. A crucial ingredient in the proof is that we assume below (without loss of generality) that all pebblings are *frugal*, meaning that no obviously redundant pebble placements are made, but that all pebbles placed on the graph are used to place other black pebbles on successors or to remove white pebbles from successors. (Again, we refer to Section 3 for a more thorough discussion of these pebbling technicalities.)

**Lemma 2.7.** *Suppose that $\mathcal{P} = \{\mathbb{P}_\sigma, \ldots, \mathbb{P}_\tau\}$ is a conditional black-white pebbling on $\Gamma(c, r)$ such that*

1. $\max\{space(\mathbb{P}_\sigma), space(\mathbb{P}_\tau)\} < s$ *for* $0 < s \leq c/8 - 1$.

2. $\mathcal{P}$ *pebbles all sinks in $\Gamma(c, r)$ during the time interval* $[\sigma, \tau]$.

3. $space(\mathcal{P}) < r + s + 2$.

*Then it holds that* $time(\mathcal{P}) = \tau - \sigma \geq \left(\frac{c-2s}{4s+4}\right)^r \cdot r!$.

To establish this result we will need the following four technical lemmas, the proofs of which are postponed to Section 6. Lemmas 2.8 and 2.9 are easy, but Lemmas 2.10 and 2.11 are somewhat less immediate and provide the key to the proof.

---

[4]The formal definition will be given later in Definition 3.4, but as an example the graph in Figure 1(a) is a pyramid of height 2.

**Lemma 2.8.** *Suppose $v$ is a vertex with a path $Q$ to some sink such that all vertices in $Q$ have outdegree 1. Then any frugal black-white pebbling pebbles $v$ exactly once, and the path $Q$ contains pebbles during one contiguous time interval.*

**Lemma 2.9.** *Let $H$ be a subgraph of $G$ such that the only edges between $V(H)$ and $V(G) \setminus V(H)$ emanate from the unique sink $z_h$ of $H$. Suppose that $\mathcal{P}$ is a complete pebbling of $G$ such that $H$ is completely empty of pebbles at some time $\tau'$ but at least one vertex of $H$ has been pebbled during the time interval $[0, \tau']$. Then $\mathcal{P}$ must have pebbled $H$ completely during the interval $[0, \tau']$.*

**Lemma 2.10.** *At all times during a pebbling of $\Gamma(c, r)$ as in Lemma 2.7, strictly less than $4(s+1)$ pyramids $\Pi_{2r}^{(j)}$ contain pebbles simultaneously.*

**Lemma 2.11.** *At all times during a pebbling of $\Gamma(c, r)$ as in Lemma 2.7, strictly less than $4(s+1)$ spine sections contain pebbles simultaneously.*

*Proof of Lemma 2.7.* Let $\mathcal{P} = \{\mathbb{P}_\sigma, \ldots, \mathbb{P}_\tau\}$ be a pebbling as in the statement of the lemma. We show that $time(\mathcal{P}) \geq T(c, r, s) = \left(\frac{c-2s}{4s+4}\right)^r \cdot r!$ by induction over $r$.

For $r = 1$, the assumptions in the lemma imply that more than $c - 2s$ sinks are empty at times $\sigma$ and $\tau$. These sinks must be pebbled, which trivially requires strictly more than $c - 2s > \left(\frac{c-2s}{4s+4}\right) = T(c, 1, s)$ time steps.

Assume that the lemma holds for $\Gamma(c, r-1)$ and consider any pebbling of $\Gamma(c, r)$. Less than $2s$ spines contain pebbles at time $\sigma$ or time $\tau$. All the other strictly more than $c - 2s$ spines are empty at times $\sigma$ and $\tau$ but must be completely pebbled during $[\sigma, \tau]$ since their sinks are pebbled during this time interval. (This can be more formally argued by using Lemma 3.12.)

Consider the first time $\sigma'$ when any spine gets a pebble for the first time. Let us denote this spine by $Q'$. By Lemma 2.8 we know that $Q'$ contains pebbles during a contiguous time interval until it is completely pebbled and emptied at, say, time $\tau'$. During this whole interval $[\sigma', \tau']$ less than $4s + 4$ sections contain pebbles at any one given time by Lemma 2.11, so in particular less than $4s + 4$ spines contain pebbles. Moreover, Lemma 2.8 says that every spine containing pebbles will remain pebbled until completed. What this means is that if we order the spines with respect to the time when they first receive a pebble in groups of size $4s + 4$, no spine in the second group can be pebbled until the at least one spine in the first group has been completed.

We observe that this divides the spines that are empty at the beginning and end of $\mathcal{P}$ into strictly more than $\frac{c-2s}{4s+4}$ groups. Furthermore, we claim that completely pebbling just one empty spine requires at least $r \cdot T(c, r-1, s)$ time steps. Given this claim we are done, since it follows that the total pebbling time must then be lower-bounded by $\frac{c-2s}{4s+4} r \cdot T(c, r-1, s) = T(c, r, s)$. This is so since at least one spine from each group is pebbled in a time interval totally disjoint from the time intervals for all spines in the next group.

It remains to establish the claim. To this end, fix any spine $Q^*$ empty at times $\sigma^*$ and $\tau^*$ but completely pebbled in $[\sigma^*, \tau^*]$. Consider the first time $\tau_1 \in [\sigma^*, \tau^*]$ when any section in $Q^*$, let us denote it by $R_1$, has been completely pebbled (i.e., all vertices has been touched by pebbles but are now empty again). During the time interval $[\sigma^*, \tau_1]$ all pyramid sinks $z_1, \ldots, z_c$ must be pebbled (since they are immediate predecessors). Since less than $2 \cdot (4s + 4) < c$ pyramids contain pebbles at times $\sigma^*$ or $\tau_1$ (Lemma 2.10), at least one pyramid is pebbled completely (Lemma 2.9), which requires $r + 1$ pebbles. Moreover, there is at least one pebble on the section $R_1$ during this whole interval. Hence, there must exist a point in time $\sigma_1 \in [\sigma^*, \tau_1]$ when there are strictly less than $(r + 2) + s - (r + 1) - 1 = s$ pebbles on the subgraph $\Gamma(c, r - 1)$. Also, at this time $\sigma_1$ less than $4s + 4$ sections contain pebbles (Lemma 2.11), and in particular this means that there are pebbles on less than $4s + 3$ other section of our spine $Q^*$. This puts an upper bound on the number of sections of $Q^*$ that can have been touched by pebbles this far, since every section is completely pebbled during a contiguous time interval before being emptied again, and we chose to focus on the first section $R_1$ in $Q^*$ that was finished.

Look now at the first section $R_2$ in $Q^*$ *other than the less than* $4s + 4$ *sections containing pebbles at time* $\sigma_1$ that is completely pebbled, and let the time when $R_2$ is finished be denoted $\tau_2$ (clearly, $\tau_2 > \tau_1$). During $[\sigma_1, \tau_2]$ all sinks of $\Gamma(c, r-1)$ must have been pebbled, and at time $\tau_2 - 1$ less than $4s + 3$ other section in $Q^*$ contain pebbles.

Finally, consider the first new section $R_3$ in our spine $Q^*$ to be completely pebbled among those not yet touched at time $\tau_2 - 1$. Suppose that $R_3$ is finished at time $\tau_3$. Then during $[\tau_2, \tau_3]$ some pyramid is completely pebbled, and thus there is some time $\sigma_3 \in (\tau_2, \tau_3)$ when there are at least $r + 1$ pebbles on this pyramid and at least one pebble on the spine $Q^*$, leaving less than $s$ pebbles for $\Gamma(c, r-1)$. But this means that we can apply the induction hypothesis on the interval $[\sigma_1, \sigma_3]$ and deduce that $\sigma_3 - \sigma_1 \geq T(c, r-1, s)$. Note also that at time $\sigma_3$ less than $8s + 8 < c$ sections in $Q^*$ have been finished.

Continuing in this way, for every group of $8s + 8 < c$ finished sections in the spine $Q^*$ we get one pebbling of $\Gamma(c, r-1)$ in space less than $r + s + 1$ and with less than $s$ pebbles in the start and end configurations, which allows us to apply the induction hypothesis a total number of at least $\frac{cr}{8s+8} > r$ times. (Just to argue that we get the constants right, note that $8s + 8 < c$ implies that after the final pebbling of the sinks of $\Gamma(c, r-1)$ has been done, there is still some empty section left in $Q^*$. When this final section is taken care of, we will again get at least $r + 1$ pebbles on some pyramid while at least one pebble resides on $Q^*$, so we get the space on $\Gamma(c, r-1)$ down below $s$ as is needed for the induction hypothesis.)

This proves our claim that pebbling one spine takes time at least $r \cdot T(c, r-1, s)$. Lemma 2.7 now follows. □

# 3   Preliminaries

In this section, we collect all the basic definitions and facts we need about resolution and pebbling.

## 3.1   The Resolution Proof System

A *literal* is either a propositional logic variable or its negation, denoted $x$ and $\overline{x}$, respectively. A *clause* $C = a_1 \vee \cdots \vee a_k$ is a set of literals. A clause containing at most $k$ literals is called a $k$-*clause*. A *CNF formula* $F = C_1 \wedge \cdots \wedge C_m$ is a set of clauses. A $k$-*CNF formula* is a CNF formula consisting of $k$-clauses. We say that $F$ *implies* $C$, denoted $F \vDash C$, if any truth value assignment satisfying $F$ must also satisfy $C$.

When we want to study length and space simultaneously, the following definition of the resolution proof system is very convenient.

**Definition 3.1 (Resolution ([ABRW02])).** A sequence of *clause configurations* (sets of clauses) $\pi = \{\mathbb{C}_0, \ldots, \mathbb{C}_\tau\}$ is a *resolution refutation* of a CNF formula $F$ if $\mathbb{C}_0 = \emptyset$, $\mathbb{C}_\tau$ contains the contradictory *empty clause* $0$ without any literals, and for all $t \in [\tau]$, $\mathbb{C}_t$ is obtained from $\mathbb{C}_{t-1}$ by one of the following rules:

**Axiom Download** $\mathbb{C}_t = \mathbb{C}_{t-1} \cup \{C\}$ for some $C \in F$ (an *axiom* clause).

**Erasure** $\mathbb{C}_t = \mathbb{C}_{t-1} \setminus \{C\}$ for some $C \in \mathbb{C}_{t-1}$.

**Inference** $\mathbb{C}_t = \mathbb{C}_{t-1} \cup \{D\}$ for some $D$ inferred from $C_1, C_2 \in \mathbb{C}_{t-1}$ by the *resolution rule*, i.e., $D = C_1 \cup C_2 \setminus \{x, \overline{x}\}$ for some variable $x$ such that $x \in C_1$ and $\overline{x} \in C_2$.

**Definition 3.2 (Length and space).** The *length* $L(\pi)$ of a resolution derivation $\pi$ is the total number of axiom downloads and inferences made in $\pi$, i.e., the total number of clauses counted with repetitions.

The *clause space* $Sp(\mathbb{C})$ of a clause configuration $\mathbb{C}$ is $|\mathbb{C}|$, i.e., the number of clauses in $\mathbb{C}$, and the *total space* $TotSp(\mathbb{C})$ is $\sum_{C \in \mathbb{C}} |C|$, i.e., the total number of literals in $\mathbb{C}$ counted with repetitions. The clause space (total space) of a derivation $\pi$ is the maximal clause space (total space) of any clause configuration $\mathbb{C} \in \pi$.

Taking the minimum over all refutations of a formula $F$, we define $L(F \vdash 0) = \min_{\pi : F \vdash 0}\{L(\pi)\}$, $Sp(F \vdash 0) = \min_{\pi : F \vdash 0}\{Sp(\pi)\}$, and $TotSp(F \vdash 0) = \min_{\pi : F \vdash 0}\{TotSp(\pi)\}$ as the length, clause space, and total space of refuting $F$ in resolution, respectively.

It is sometimes technically convenient to add a *weakening* rule to Definition 3.1, allowing a resolution derivation to derive a weaker clause $C' \supsetneqq C$ from $C$. We can allow or disallow this rule as we see fit, since any such weakening steps can always be eliminated without increasing the length or space of a refutation. In particular, the following upper bounds on resolution length and space are cleaner to state if we assume that weakening can be used.

**Proposition 3.3.** *Suppose $\mathbb{C}$ is a set of clauses and $C$ is a clause, both over a set of variables of size $n$. Then $\mathbb{C} \vDash C$ if and only if there exists a resolution derivation of $C$ from $\mathbb{C}$. Furthermore, if $C$ can be derived from $\mathbb{C}$ then it can be derived in length at most $2^{n+1} - 1$ and total space at most $n(n+2)$ simultaneously.*

The proof of this proposition is standard and can be found in, for instance, [BN09b].

## 3.2 Graph Terminology and Notation

We write $G$ to denote a graph with vertices $V(G)$ and edges $E(G)$. All graphs in this paper are directed unless otherwise stated, and $(u, v)$ denotes a directed edge from $u$ to $v$.

We let $succ(v)$ denote the immediate successors and $pred(v)$ denote the immediate predecessors of a vertex $v$ in $G$. We say that vertices of $G$ with indegree 0 are *sources* and that vertices with outdegree 0 are *sinks*. (In the literature, sources are also referred to as *inputs* and sinks as *targets* or *outputs*). In the notation just introduced, a source vertex $s$ in $G$ is a vertex with $pred(s) = \emptyset$, and for a sink $z$ we have $succ(z) = \emptyset$. We will write $S(G)$ to denote the source vertices of $G$ and $Z(G)$ to denote the sink vertices. For brevity, we will sometimes refer to a DAG with a unique sink as a *single-sink DAG*.

Some more notational conventions are that the parameter $\ell$ denotes the maximal indegree of a DAG, and that when not stated otherwise, $n$ will denote the size, i.e., the number of vertices, of a DAG (or, if more convenient, the size to within a small constant factor). We write $Q : v \rightsquigarrow w$ to denote a path $Q$ starting at the vertex $v$ and ending at the vertex $w$.

The *pyramid graphs* already mentioned several times in this paper are formally defined as follows.

**Definition 3.4 (Pyramid graph).** The *pyramid graph* $\Pi_h$ of height $h$ is a layered DAG with $h + 1$ levels, where there is one vertex on the highest level (the sink $z$), two vertices on the next level et cetera down to $h + 1$ vertices at the lowest level 0. The $i$th vertex at level $L$ has incoming edges from the $i$th and $(i + 1)$st vertices at level $L - 1$.

## 3.3 Pebbling Technicalities

The flavour of the pebble game presented in Definition 1.1 is the version that we are interested in for our applications in proof complexity, but for the purposes of stating and proving our results we need a slightly more general definition.

**Definition 3.5 (General pebbling definition).** Suppose that $G$ is a DAG with sources $S$ and sinks $Z$ (one or many). A *black-white pebbling* from $(B_0, W_0)$ to $(B_\tau, W_\tau)$ in $G$ is a sequence of pebble configurations $\mathcal{P} = \{\mathbb{P}_0, \ldots, \mathbb{P}_\tau\}$ such that $\mathbb{P}_0 = (B_0, W_0)$, $\mathbb{P}_\tau = (B_\tau, W_\tau)$, and for all $t \in [\tau]$, $\mathbb{P}_t$ follows from $\mathbb{P}_{t-1}$ by one of the rules in Definition 1.1. The space of a pebble configuration $\mathbb{P} = (B, W)$ is $\boldsymbol{space}(\mathbb{P}) = |B \cup W|$ and the space of the pebbling $\mathcal{P}$ is $\boldsymbol{space}(\mathcal{P}) = \max_{t \in [\tau]}\{\boldsymbol{space}(\mathbb{P}_t)\}$.

We say that a pebbling $\mathcal{P} = \{\mathbb{P}_0, \ldots, \mathbb{P}_\tau\}$ is *conditional* if $\mathbb{P}_0 \neq (\emptyset, \emptyset)$ and *unconditional* otherwise.

A complete black-white pebbling *visiting* $Z$ is a pebbling such that $\mathbb{P}_0 = \mathbb{P}_\tau = (\emptyset, \emptyset)$ and such that for every $z \in Z$, there exists a time $t_z \in [\tau]$ when $z \in B_{t_z} \cup W_{t_z}$. The minimum space of such a visiting pebbling is denoted $\textit{BW-Peb}^\emptyset(G)$, and for the black pebble game we have the measure $\textit{Peb}^\emptyset(G)$.

A *persistent* pebbling of $G$ is a pebbling $\mathcal{P}$ such that $\mathbb{P}_\tau = (Z, \emptyset)$. The minimum space of any complete persistent black-white or black-only pebbling of $G$ is denoted $\textit{BW-Peb}(G)$ and $\textit{Peb}(G)$, respectively.

We think of the moves in a pebbling as occurring at integral time intervals $t = 1, 2, \ldots$ and talk about the pebbling move "at time $t$" (which is the move resulting in configuration $\mathbb{P}_t$) or the moves "during the time interval $[t_1, t_2]$."

A visiting pebbling touches all sinks but leaves the graph empty at time $\tau$, whereas a persistent pebbling leaves black pebbles on all sinks at the end of the pebbling. If $G$ has $m$ sinks, then it clearly holds that $\textit{BW-Peb}(G) \leq \textit{BW-Peb}^\emptyset(G) + m$ and $\textit{Peb}(G) \leq \textit{Peb}^\emptyset(G) + m$. Also, if $G$ has a unique sink, it is easy to see that $\textit{Peb}(G) = \textit{Peb}^\emptyset(G)$.

The only pebblings we are really interested in are complete pebblings of $G$. However, when we prove lower bounds on pebbling price it will sometimes be convenient to be able to reason in terms of partial pebbling move sequences, i.e., conditional pebblings. One can think of conditional pebblings as pebblings that receive the start configuration $(B_1, W_1)$ "as a gift", and are also allowed to leave $(B_2, W_2)$ without "cleaning up" when they finish. It is clear that we can assume that $(B_1, W_1) = (B_1, \emptyset)$ and $(B_2, W_2) = (\emptyset, W_2)$ since we can freely place white pebbles on $G$ and freely remove black pebbles. The way the gift can help us is that we get black pebbles at the beginning for free, and are allowed to leave white pebbles without having to do the hard work of removing them.

The reason we need visiting pebblings and not just persistent ones is that the graphs of interest will be constructed in terms of smaller subgraph components with useful pebbling properties, and that for such subgraphs we have the following easy observation (the proof of which is omitted).

**Observation 3.6.** *Suppose that $G$ is a DAG and that $\mathcal{P}$ is any complete pebbling of $G$. Let $U \subseteq V(G)$ be any subset of vertices of $G$ and let $H = H(G, U)$ denote the induced subgraph with vertices $V(H) = U$ and edges $E(H) = \{(u, v) \in E(G) \,|\, u, v \in U\}$. Then the pebbling $\mathcal{P}$ restricted to the vertices in $U$ is a complete visiting pebbling strategy for $H$.*

Some proofs are facilitated by observing that visiting pebblings have a certain "duality" property. The next proposition is an immediate consequence of the anti-symmetric nature of the pebbling rules in Definition 1.1 (just observe that the rules for placing and removing a black pebble are the duals of the rules for removing and placing a white pebble, respectively).

**Proposition 3.7 ([CS76]).** *If $\mathcal{P}$ is a black-white pebbling from $(B_1, W_1)$ to $(B_2, W_2)$, then we can get a dual pebbling $\overline{\mathcal{P}}$ from $(W_2, B_2)$ to $(W_1, B_1)$ in exactly the same time and space by reversing the sequence of moves and switching the colours of the pebbles. In particular, if $\mathcal{P}$ is a complete visiting pebbling of $G$, then so is $\overline{\mathcal{P}}$.*

For the applications in proof complexity, we often want results stated for DAGs with one unique sink, but most pebbling results are more natural to state and prove for DAGs with multiple sinks. This small technicality is easily taken care of as follows.

**Definition 3.8 (Single-sink version).** Let $G$ be a DAG with sinks $Z(G) = \{z_1, \ldots, z_m\}$ for $m > 1$. The *single-sink version* $\widehat{G}$ of $G$ consists of all vertices and edges in $G$ plus the extra vertices $z_1^*, \ldots, z_{m-1}^*$ and the edges $(z_1, z_1^*)$, $(z_2, z_1^*)$, $(z_1^*, z_2^*)$, $(z_3, z_2^*)$, $(z_2^*, z_3^*)$, $(z_4, z_3^*)$, et cetera up to $(z_{m-2}^*, z_{m-1}^*)$, $(z_m, z_{m-1}^*)$.

That is, $\widehat{G}$ consists of $G$ with a binary tree of minimal size added on top of the sinks. See Figure 5 for a small example. The following observation is immediate.

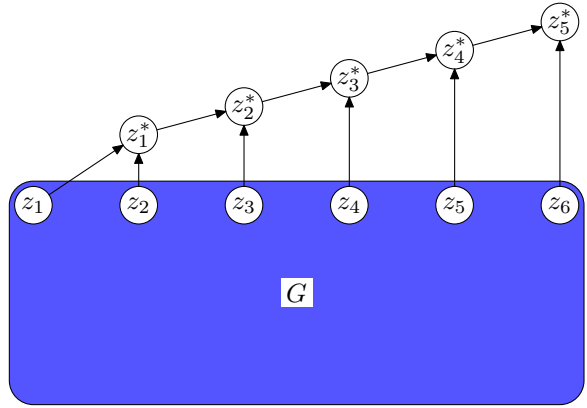

**Figure 5:** Schematic illustration of single-sink version $\widehat{G}$ of graph $G$.

**Observation 3.9.** *Let $G$ be a DAG with sinks $Z(G) = \{z_1, \ldots, z_m\}$ for $m > 1$. Then for any flavour of pebbling (visiting or persistent) it holds that $\mathsf{BW\text{-}Peb}(\widehat{G}) \leq \mathsf{BW\text{-}Peb}(G) + 1$ and $\mathsf{Peb}(\widehat{G}) \leq \mathsf{Peb}(G) + 1$. Moreover, if there is a pebbling strategy $\mathcal{P}$ (visiting or persistent) for $G$ that can pebble the sinks in arbitrary order, then there is a pebbling strategy $\widehat{\mathcal{P}}$ of the same type (black or black-white, visiting or persistent) for $\widehat{G}$ with $\mathsf{time}(\widehat{\mathcal{P}}) \leq \mathsf{time}(\mathcal{P}) + 2m$ and $\mathsf{space}(\widehat{\mathcal{P}}) \leq \mathsf{space}(\mathcal{P}) + 1$.*

The next proposition is convenient when composing pebblings of smaller subgraphs into a pebbling of a larger graph.

**Proposition 3.10.** *Suppose that $G$ is a DAG with unique sink $z$. Then for any complete black or black-white pebbling $\mathcal{P}$ of $G$ there is a complete pebbling $\mathcal{P}'$ with the same colours such that $\mathsf{time}(\mathcal{P}') = \mathsf{time}(\mathcal{P})$, $\mathsf{space}(\mathcal{P}') = \mathsf{space}(\mathcal{P})$, and there is a time $t$ during $\mathcal{P}'$ when $z$ has a pebble but the pebbling space is strictly less than $\mathsf{space}(\mathcal{P})$.*

*Proof.* For black pebblings this statement is obvious. Once we place a black pebble on the sink $z$, we can remove all other pebbles from the DAG.

Suppose for a black-white pebbling $\mathcal{P}$ that the pebbling space reaches the maximum $s$ precisely when a pebble is placed on $z$ at time $t$. Then the move at time $t + 1$ must be a pebble removal. If a pebble is removed from a vertex other than $z$, we are done. Otherwise, fix some vertex $w \in pred(z)$ having $z$ as its only successor. Suppose that $w$ contains a white pebble during some interval $[\sigma, \tau] \supseteq [t, t + 1]$ (and if not, run the dual pebbling in Proposition 3.7 instead). To obtain $\mathcal{P}'$, we change $\mathcal{P}$ as follows. The pebble placement on $w$ at time $\sigma$ is omitted. At time $t$, a white pebble is placed on $z$. In between times $t$ and $t + 1$, $w$ is white pebbled, and then the white pebble on $z$ is removed at time $t + 1$. $\qquad\square$

It is immediate from the definition of the black pebble game that black pebblings always proceed in a bottom-up fashion in the following sense.

**Observation 3.11.** *Suppose that $Q : u \rightsquigarrow v$ is a path in $G$ and that $\mathcal{P} = \{\mathbb{P}_\sigma, \mathbb{P}_{\sigma+1}, \ldots, \mathbb{P}_\tau\}$ is a black-only pebbling such that the whole path $Q$ is completely free of pebbles at time $\sigma$ but a pebble is placed on the endpoint $v$ at time $\tau$. Then the starting point $u$ must have been pebbled during the time interval $(\sigma, \tau)$.*

A simple but important lemma, lying at the heart of essentially all black-white pebbling lower bounds, is the following generalization of Observation 3.11 to black-white pebbling: In order to pebble the endpoint $v$ of a some path, one needs to pebble all vertices on this path at some point prior to *or after* pebbling $v$.

**Lemma 3.12 ([GT78]).** *Suppose that $Q : u \rightsquigarrow v$ is a path in $G$ and that $\mathcal{P} = \{\mathbb{P}_\sigma, \mathbb{P}_{\sigma+1}, \ldots, \mathbb{P}_\tau\}$ is a black-white pebbling such that the whole path $Q$ is completely free of pebbles at times $\sigma$ and $\tau$ but the endpoint $v$ is pebbled at some point during $(\sigma, \tau)$. Then the starting point $u$ is pebbled during $(\sigma, \tau)$ as well.*

*Proof.* By induction over the length of the path $Q$. The base case $u = v$ is trivial. For the induction step, let $w$ be the immediate successor of $u$ on $Q$. By the induction hypothesis, $w$ is pebbled and unpebbled again some time during $(\sigma, \tau)$. Then $u$ must be covered by a pebble either when the pebble on $w$ is placed there (if this pebble is black) or when it is removed (if it is white). The lemma follows. □

When proving lower bounds on pebblings, it often helps to assume that the pebblings under consideration do not perform any obviously redundant moves. The following definition, which formalizes this notion, is a generalization of [GLT80] from black-only to black-white pebbling.

**Definition 3.13 (Frugal pebbling).** Let $\mathcal{P}$ be a complete pebbling of a DAG $G$. To every pebble placement on a vertex $v$ at time $\sigma$ we associate the *pebbling interval* $[\sigma, \tau)$, where $\tau = \tau(\sigma, v)$ is the first time after $\sigma$ when the pebble is removed from $v$ again (or $\tau = \infty$, say, if this never happens).

If a sink $z_i \in Z(G)$ is pebbled for the first time at time $\sigma$, then the pebble on $z_i$ is *essential* during the pebbling interval $[\sigma, \tau)$. A pebble on a non-sink vertex $v$ is essential during $[\sigma, \tau)$ if either an essential black pebble is placed on an immediate successor of $v$ during $(\sigma, \tau)$ or an essential white pebble is removed from an immediate successor of $v$ during $(\sigma, \tau)$. Any other pebble placements on any vertices are non-essential.

The pebbling strategy $\mathcal{P}$ is *frugal* if all pebbles in $\mathcal{P}$ are essential at all times.

Without loss of generality, we can assume that all pebblings are frugal.

**Lemma 3.14.** *For any complete pebbling $\mathcal{P}$ (black or black-white, visiting or persistent) there is a frugal pebbling $\mathcal{P}'$ of the same type such that $\mathsf{time}(\mathcal{P}') \leq \mathsf{time}(\mathcal{P})$ and $\mathsf{space}(\mathcal{P}') \leq \mathsf{space}(\mathcal{P})$.*

*Proof sketch.* Just delete any non-essential pebbles and verify that what remains is a legal pebbling. □

One minor technical snag is that we will need to assume not only that complete pebblings are frugal, but that this also holds for *conditional pebblings* (Definition 3.5). This is no real problem, however, since we can always assume that the conditional pebblings we are dealing with are subpebblings of larger, unconditional pebblings. In fact, an alternative way of defining frugal pebblings (unconditional or conditional) is to say that a pebble placement on a non-sink vertex $v$ is essential if the pebble stays until either a black pebble is placed on an immediate successor of $v$ or a white pebble is removed from an immediate successor of $v$. If a pebbling contains non-essential moves, then it is easy to see that such moves can be eliminated to get a shorter pebbling that is still legal. This new pebbling might contain other non-essential moves, but after applying the procedure a finite number of times we obtain a pebbling with only essential moves. Adding the requirement that each sink should only be pebbled once, we recover Definition 3.13.

We conclude this section by recalling the following results for pebblings of pyramid graphs.

**Theorem 3.15 ([Coo74, Kla85]).** *The black pebbling price of the pyramid $\Pi_h$ of height $h$ is $\mathsf{Peb}(\Pi_h) = h + 2$, and there is a linear-time pebbling achieving this bound.*

*The black-white pebbling price of $\Pi_h$ is $\mathsf{BW\text{-}Peb}(\Pi_h) = h/2 + \mathrm{O}(1)$. For pyramids of odd height the exact bound $\mathsf{BW\text{-}Peb}(\Pi_{2h+1}) = h + 3$ holds, and for even height we have $\mathsf{BW\text{-}Peb}^\emptyset(\Pi_{2h}) = h + 2$.*

We remark that the exact bounds for black-white pebbling above are not stated or proven by Klawe in [Kla85], but can be read off from the exposition of Klawe's proof in (the full-length version [NH08a] of) [NH08b].

# 4    Labelled Black-White Pebblings and Resolution Simulations

Let us now prove the claims made in Section 2.1 about the labelled black-white pebble game in Definition 2.2, namely that this game is just a limited version of standard black-white pebbling (Lemma 2.3) and that resolution refutations of pebbling contradictions can simulate labelled pebblings if all labelled pebble subconfigurations have bounded size (Lemma 2.5).

## 4.1    Proof of Lemma 2.3

Recall that we want to prove that if $\mathcal{L}$ is a complete labelled pebbling of a single-sink DAG $G$, then we can transform $\mathcal{L}$ into a complete standard black-white pebbling $\mathcal{P}_{\mathcal{L}}$ of $G$ with $\textit{time}(\mathcal{P}_{\mathcal{L}}) \leq \frac{4}{3}\textit{time}(\mathcal{L})$ and $\textit{space}(\mathcal{P}_{\mathcal{L}}) \leq \textit{space}(\mathcal{L})$. The proof of this fact is not hard, and much of the needed material can be extracted from similar arguments in [Nor09]. Since what is actually proven in [Nor09] is something different and slightly weaker, however, we provide a full, explicit proof of Lemma 2.3 below.

The first modification of the pebble game when going from Definition 1.1 to Definition 2.2 is that in the context of resolution, a more natural rule for white pebble removal appears to be that a white pebble can be removed from a vertex when a black pebble is placed on that same vertex. It seems intuitively fairly obvious that this rule change should not really affect the pebble game, and indeed it does not.

**Lemma 4.1.** *Let us say that a* superpositioned *black-white pebbling of $G$ is a pebbling as in Definition 1.1, except that a vertex may have both a black and a white pebble on itself, and that rule (4) is changed to:*

    *4'.  A white pebble on $v$ can be removed only if there is a black pebble on $v$.*

*Then for any complete superpositioned pebbling $\mathcal{S}$ of $G$ there is a standard complete black-white pebbling $\mathcal{P}$ with $\textit{time}(\mathcal{P}) \leq \textit{time}(\mathcal{S})$ and $\textit{space}(\mathcal{P}) \leq \textit{space}(\mathcal{S})$.*

*Proof.* Suppose that we are given a superpositioned pebbling $\mathcal{S} = \{\mathbb{S}_0, \dots, \mathbb{S}_\tau\}$ of $G$. We construct a standard black-white pebbling $\mathcal{P} = \{\mathbb{P}_0, \dots, \mathbb{P}_\tau\}$ such that for $\mathbb{P}_t = (B_t, W_t)$ and $\mathbb{S}_t = (B'_t, W'_t)$ it holds that $B_t \supseteq B'_t$, $B_t \cup W_t = B'_t \cup W'_t$ and (as required by Definition 1.1) $B_t \cap W_t = \emptyset$. In particular, this means that $\textit{space}(\mathcal{P}) = \textit{space}(\mathcal{S})$, and that if $\mathcal{S}$ is a complete pebbling, then so is $\mathcal{P}$.

The construction is by forward induction over $\mathcal{S}$. We set $\mathbb{P}_0 = \mathbb{S}_0 = (\emptyset, \emptyset)$ and then make the inductive step by a case analysis over the pebbling moves.

1. If $\mathcal{S}$ places a black pebble on $v$ at time $t+1$, the vertices in $pred(v)$ must be pebbled in $\mathbb{S}_t$ and thus by induction also in $\mathbb{P}_t$. If $v \in W_t$, we remove the white pebble from $v$ in $\mathcal{P}$. Then we place a black pebble on $v$.

2. If $\mathcal{S}$ removes a black pebble from $v$ at time $t+1$, by induction $v$ is black-pebbled in $\mathbb{P}_t$. We remove the black pebble from $v$ in $\mathcal{P}$, unless $v \in W'_t$ in which case we leave the black pebble on $v$.

3. If $\mathcal{S}$ places a white pebble on $v$ at time $t+1$, we place a white pebble there in $\mathcal{P}$ if $v \notin B_t$ and otherwise do nothing.

4. When a white pebble is removed from $v$ in $\mathcal{S}$ it holds that $v \in B'_t$. Thus, by induction $v \in B_t$, so the white pebble has already been removed from $v$ in $\mathcal{P}$, or was never placed there.

It clearly holds that $\textit{time}(\mathcal{P}) \leq \textit{time}(\mathcal{S})$, since $\mathcal{P}$ makes at most as many pebbling moves as $\mathcal{S}$.      $\square$

The second step in the proof of Lemma 2.3 is to show that if we take a complete labelled pebbling $\mathcal{L} = \{\mathbb{L}_0, \dots, \mathbb{L}_\tau\}$ of a DAG $G$ and look at the vertices $\big(Bl(\mathbb{L}_t), Wh(\mathbb{L}_t)\big)$ covered by black and white pebbles for all $t \in [\tau]$, we can extract a legal complete (superpositioned) black-white pebbling of $G$ in essentially the same time and space. We prove this formally in the next two lemmas.

The first lemma says that without loss of generality we can assume that all labelled pebblings are *non-redundant* in the sense that if a subconfiguration $v\langle V \rangle$ is derived at time $t$, then this subconfiguration is not just thrown away but is used at some time $t' > t$ further on in the pebbling before being erased.

**Lemma 4.2.** *Let $\mathcal{L} = \{\mathbb{L}_0, \ldots, \mathbb{L}_\tau\}$ be any complete labelled pebbling of a DAG $G$. Then we can construct a complete labelled pebbling $\mathcal{L}' = \{\mathbb{L}'_0, \ldots, \mathbb{L}'_{\tau'}\}$ of $G$ with $\mathsf{time}(\mathcal{L}') \leq \mathsf{time}(\mathcal{L})$ and $\mathsf{space}(\mathcal{L}') \leq \mathsf{space}(\mathcal{L})$ that has the following property: If $v\langle V \rangle$ is erased at time $t$ in $\mathcal{L}'$, i.e., $v\langle V \rangle \in \mathbb{L}'_t \setminus \mathbb{L}'_{t+1}$, then this subconfiguration has been used in a merger or reversal move immediately before being erased, and the subconfiguration resulting from this move is present in $\mathbb{L}'_{t+1}$.*

*Proof.* This is easy if formally somewhat tedious, so let us first try to visualize the proof. For any labelled pebbling $\mathcal{L}$, we can construct a DAG $G_\mathcal{L}$ encoding the pebbling as follows. For every subconfiguration $v\langle V \rangle$ appearing at time $t_1$ and staying in the graph until time $t_2$ when it is erased, we create a vertex $(v\langle V \rangle, [t_1, t_2])$. For each merger $u\langle U \rangle = \mathsf{merge}(v\langle V \rangle, w\langle W \rangle)$, we draw edges from $v\langle V \rangle$ and $w\langle W \rangle$ to $u\langle U \rangle$. The sources in $G_\mathcal{L}$ are vertices $(v\langle pred(v) \rangle, [t_1, t_2])$, and by assumption there is a sink $(z\langle \emptyset \rangle, [t_1, \tau])$. Note that without loss of generality we can assume that we never derive a subconfiguration that is already present in the graph, so all vertices in $G_\mathcal{L}$ have indegree 0 or 2 corresponding to introductions and mergers, respectively.

Consider the subgraph of $G_\mathcal{L}$ consisting of all vertices from which the sink vertex $(z\langle \emptyset \rangle, [t_1, \tau])$ is reachable. We construct $\mathcal{L}'$ to be the subpebbling corresponding exactly to the moves in this subgraph, except that we reorder moves if needed so that erasures are always performed as soon as possible. Since the moves in $\mathcal{L}'$ are a subset of the moves in $\mathcal{L}$, clearly $\mathsf{time}(\mathcal{L}') \leq \mathsf{time}(\mathcal{L})$.

Formally, this amounts to the following. We construct the modified pebbling $\mathcal{L}'$ by backward induction over $\mathcal{L} = \{\mathbb{L}_0, \ldots, \mathbb{L}_\tau\}$. Let $\mathbb{L}'_\tau = \mathbb{L}_\tau = \{z\langle \emptyset \rangle\}$. Our induction hypothesis is that $\mathbb{L}'_{t^*} \subseteq \mathbb{L}_{t^*}$ for $t^* > t$. The backward induction step from $t+1$ to $t$ is a case analysis over the moves $\mathbb{L}_t \rightsquigarrow \mathbb{L}_{t+1}$ in $\mathcal{L}$. For simplicity, we allow using fractional time steps in the interval $[t, t+1]$ in the inductive constructions below.

**Introduction** $\mathbb{L}_{t+1} = \mathbb{L}_t \cup \{v\langle pred(v) \rangle\}$: Set $\mathbb{L}'_t = \mathbb{L}'_{t+1} \setminus \{v\langle pred(v) \rangle\}$. Note that we might have $\mathbb{L}'_t = \mathbb{L}'_{t+1}$ if $v\langle pred(v) \rangle \notin \mathbb{L}'_{t+1}$. In any case, the induction hypothesis holds for $\mathbb{L}'_t$.

**Merger** $\mathbb{L}_{t+1} = \mathbb{L}_t \cup \{v\langle (V \cup W) \setminus \{w\} \rangle\}$: If $v\langle (V \cup W) \setminus \{w\} \rangle \notin \mathbb{L}'_{t+1}$, set $\mathbb{L}'_t = \mathbb{L}'_{t+1}$. The induction hypothesis trivially remains true. Otherwise, if the merged subconfiguration is present in $\mathbb{L}'_{t+1}$ set $\mathbb{L}'_t = (\mathbb{L}'_{t+1} \cup \{v\langle V \rangle, w\langle W \rangle\}) \setminus \{v\langle (V \cup W) \setminus \{w\} \rangle\}$. We can go from $\mathbb{L}'_t$ to $\mathbb{L}'_{t+1}$ in at most three steps via intermediate L-configurations $\mathbb{L}'_{t+1/3} = \mathbb{L}'_t \cup \{v\langle (V \cup W) \setminus \{w\} \rangle\}$ and $\mathbb{L}'_{t+2/3} = \mathbb{L}'_{t+1} \cup \{w\langle W \rangle\}$ by first merging $v\langle V \rangle$ and $w\langle W \rangle$, then possibly erasing $v\langle V \rangle$ and finally possibly erasing $w\langle W \rangle$.

**Erasure** $\mathbb{L}_{t+1} = \mathbb{L}_t \setminus \{v\langle V \rangle\}$: All erasure moves in $\mathcal{L}'$ are taken care of in connection with mergers, so set $\mathbb{L}'_t = \mathbb{L}'_{t+1}$.

We claim that all moves in $\mathcal{L}'$ constructed in this way are legal. For if $u\langle U \rangle \in \mathbb{L}'_t$, then $u\langle U \rangle \in \mathbb{L}_t$ and we know that this subconfiguration must have been derived at some point in time $t^* \leq t$ in $\mathcal{L}$. Thus the backward construction of $\mathcal{L}'$ will yield a correct derivation of $u\langle U \rangle$. Also note that by construction, when a subconfiguration in $\mathcal{L}'$ is erased it has just been used in some merger move.

Finally, by construction $\mathbb{L}'_t \subseteq \mathbb{L}_t$, and for the intermediate fractional time step L-configurations $\mathbb{L}'_{t+a/b}$ in the merger moves in $\mathcal{L}'$ we have $\mathbb{L}'_{t+a/b} \subseteq \mathbb{L}_{t+1}$. It follows that $\mathsf{space}(\mathcal{L}') \leq \mathsf{space}(\mathcal{L})$. $\qquad\square$

For labelled pebblings as in Lemma 4.2, if we ignore all relations between black and white pebbles in the subconfigurations and consider $\big(Bl(\mathbb{L}_t), Wh(\mathbb{L}_t)\big)$ for $t \in [\tau]$, this is a legal superpositioned pebbling.

**Lemma 4.3.** *Suppose that $\mathcal{L}$ is a complete labelled pebbling of a DAG $G$. Then there is a complete superpositioned pebbling $\mathcal{S}$ of $G$ such that $\mathsf{time}(\mathcal{S}) \leq \frac{4}{3}\mathsf{time}(\mathcal{L})$ and $\mathsf{space}(\mathcal{S}) \leq \mathsf{space}(\mathcal{L})$.*

*Proof.* By Lemma 4.2, without loss of generality we can assume that each $v\langle V\rangle$ is erased from $\mathcal{L}$ precisely after it has been used in a merger, and that $v\langle V\rangle$ is erased before $w\langle W\rangle$ when both subconfigurations are eliminated after a move $v\langle(V \cup W) \setminus \{w\}\rangle = \mathsf{merge}(v\langle V\rangle, w\langle W\rangle)$, so that the white pebble on $w$ is removed before the black pebble on $w$.

It is clear that we are done if we can construct a superpositioned pebbling $\mathcal{S}$ with moves matching the moves in $\mathcal{L}$ exactly. Let $\mathbb{S}_0 = (\emptyset, \emptyset)$ and construct $\mathbb{S}_{t+1}$ inductively by looking at the moves in $\mathbb{L}_t \rightsquigarrow \mathbb{L}_{t+1}$.

*Introduction* $\mathbb{L}_{t+1} = \mathbb{L}_t \cup \{v\langle pred(v)\rangle\}$: Place white pebbles on $pred(v)$ and then a black pebble on $v$ in $\mathcal{S}$.

*Merger* $\mathbb{L}_{t+1} = \mathbb{L}_t \cup \{v\langle(V \cup W) \setminus \{w\}\rangle\}$ for $v\langle V\rangle$, $w\langle W\rangle \in \mathbb{L}_t$: No pebbling moves in $\mathcal{S}$, but note that if $v\langle V\rangle$ is now removed, the change in pebbles on $G$ in $\mathcal{L}$ is exactly the same as after an application of rule (4') on $w$.

*Erasure* $\mathbb{L}_{t+1} = \mathbb{L}_t \setminus \{v\langle V\rangle\}$: This is the only nontrivial case. In general, an erasure move in an labelled pebbling can remove an arbitrary number of white pebbles without any black pebbles being even close to these white pebbles, and there is no way we can match such a move in a superpositioned pebbling. But since we can assume that $\mathcal{L}$ is an labelled pebbling as described in Lemma 4.2, we know that $v\langle V\rangle$ has just been used in a merger. Consequently, the only pebble that disappears when going from $\big(Bl(\mathbb{L}_t), Wh(\mathbb{L}_t)\big)$ to $\big(Bl(\mathbb{L}_{t+1}), Wh(\mathbb{L}_{t+1})\big)$ is either the black pebble on $v$, which is always a legal pebble removal, or some white pebble on $w \in V$ which has just been eliminated in the merger move by a black pebble, and this is a legal pebble removal according to rule (4').

We see that $\mathcal{S}$ generated in this way is a legal superpositioned pebbling, if we modify each introduction step into $|pred(v)| + 1$ pebble placement moves. Clearly, $\mathsf{space}(\mathcal{S}) \leq \mathsf{space}(\mathcal{L})$. To see that $\mathsf{time}(\mathcal{S}) \leq \frac{4}{3}\mathsf{time}(\mathcal{L})$, consider any vertex $v$. The way $\mathcal{S}$ is constructed from $\mathcal{L}$, every time $v$ is pebbled it is both black-pebbled and white-pebbled, after which the pebbles are removed. This takes $4$ moves in $\mathcal{S}$. In $\mathcal{L}$, a single introduction move can place pebbles on many vertices. However, to remove the pebbles from $v$ requires $3$ moves, namely $1$ merger followed by $2$ erasures. This gives the time bound, and the lemma follows. $\qquad\square$

Now Lemma 2.3 follows from combining Lemmas 4.1 and 4.3.

## 4.2   Proof of Lemma 2.5

The assumption in Lemma 2.5 is that we are given a complete $(m, S)$-bounded labelled pebbling $\mathcal{L} = \{\mathbb{L}_0, \ldots, \mathbb{L}_\tau\}$ of a DAG $G$. We want to prove that for any nonconstant Boolean function $f$ of arity $d$, there is a resolution refutation $\pi_{\mathcal{L}}$ of $Peb_G[f]$ in length $L(\pi_{\mathcal{L}}) = \mathsf{time}(\mathcal{L}) \cdot \exp(\mathrm{O}(dS))$ and total space $TotSp(\pi_{\mathcal{L}}) = m \cdot \exp(\mathrm{O}(dS))$.

Let us first adopt the notation that for a vertex $v$, we let $v[f]$ denote the set of clauses obtained when substituting $f(v_1, \ldots, v_d)$ for $v$ and expanding to conjunctive normal form, and similarly for $\overline{v}[f]$. We extend this notation to clauses by defining $(C \vee D)[f] = \{C' \vee D' \mid C' \in C[f], D' \in D[f]\}$. Note that if a clause $C$ contains $K$ literals, then $C[f]$ has at most $2^{dK}$ clauses containing at most $dK$ literals each.

The proof is by induction over the pebbling $\mathcal{L}$. We maintain the invariant that if $\mathbb{L}_t$ is the set of subconfigurations at time $t$, then then $\pi$ will contain exactly the clauses $\mathbb{C}_t = \big\{\big(\bigvee_{w \in W} \overline{w} \vee v\big)[f] \mid v\langle W\rangle \in \mathbb{L}_t\big\}$. Since $\mathcal{L}$ is an $(m, S)$-bounded pebbling, this means that $\mathbb{C}_t$ will contain at most $m2^{d(1+S)}$ clauses, each clause of size at most $d(1 + S)$. To simplify the notation in the proof, we will implicitly use fractional time steps in $\pi$, making sure that it never takes more than $\exp(\mathrm{O}(dS))$ time steps to get from $\mathbb{C}_{t-1}$ to $\mathbb{C}_t$.

Consider the pebbling move made in $\mathcal{L}$ at time $t$ :

1. If $\mathcal{L}$ introduces $v\langle pred(v)\rangle$, we download all the axiom clauses in $\left(\bigvee_{w\in pred(v)} \overline{w} \vee v\right)[f]$. By assumption we have $|pred(v)| \leq S$, so the number of axiom clauses are at most $2^{d(1+S)}$.

2. Suppose $\mathcal{L}$ merges $v\langle V\rangle, w\langle W\rangle \in \mathbb{L}_{t-1}$ with $w \in V$ into $v\langle (V\cup W)\setminus\{w\}\rangle$. By the inductive hypothesis, we have the clauses $\left(\bigvee_{u\in V} \overline{u} \vee v\right)[f]$ and $\left(\bigvee_{x\in W} \overline{x} \vee w\right)[f]$ in memory. Together, these clauses clearly imply $\left(\bigvee_{u\in(V\cup W)\setminus\{w\}} \overline{u} \vee v\right)[f]$.

   Let $D$ be any clause in the set $\left(\bigvee_{u\in(V\cup W)\setminus\{w\}} \overline{u} \vee v\right)[f]$. By Proposition 3.3, we can derive $D$ from the clauses corresponding to $v\langle V\rangle$ and $w\langle W\rangle$ in length $\exp\left(\mathrm{O}(dS)\right)$ and additional total space $\mathrm{O}\left((dS)^2\right)$. Doing this in turn for all the $2^{d(1+S)}$ clauses $D \in \left(\bigvee_{u\in(V\cup W)\setminus\{w\}} \overline{u} \vee v\right)[f]$ establishes the induction step.

3. If $\mathcal{L}$ erases a subconfiguration $v\langle V\rangle$, we just erase all clauses in $\left(\bigvee_{w\in pred(v)} \overline{w} \vee v\right)[f]$ from memory.

At the end of the pebbling $\mathcal{L}$, we have $\mathbb{C}_\tau = \{z[f_d]\}$ for $z$ the sink of $G$. We conclude the refutation by downloading all the sink axioms in $\overline{z}[f_d]$ and deriving the empty clause $0$ in length $\exp(\mathrm{O}(d))$ and total space $\mathrm{O}\left(d^2\right)$. This proves the lemma.

# 5 Separations of Black Pebbling and Bounded Labelled Pebbling

The second component in our proof that resolution refutations of pebbling contradictions can be strictly more efficient than black pebblings of the corresponding graphs is to show that there are graph families which separate black pebbling and bounded black-white labelled pebbling. In this section, we briefly review the graph families exhibiting the separations between black and black-white pebbling in Lemmas 1.3, 1.5, and 1.7, and then prove that the black-white pebblings for these graphs can be carried out in the bounded labelled pebbling framework. From this Theorems 1.4, 1.6, and 1.8 immediately follow by appealing to Lemma 2.5. We first attend to Lemma 1.3 and Theorem 1.4 in Section 5.1, and then take care of Lemmas 1.5 and 1.7 and Theorems 1.6 and 1.8 in Section 5.2.

## 5.1 Bounded Pebblings for Time-Space Trade-offs

The trade-offs in Lemma 1.3 are obtained for graphs built from permutations in the following way.

**Definition 5.1 (Permutation graph ([LT82])).** Let $\pi$ denote some permutation of $\{0, 1, \dots, n-1\}$. The *permutation graph* $\Delta(n, \pi)$ over $n$ elements with respect to $\pi$ is defined as follows. $\Delta(n, \pi)$ has $2n$ vertices divided into a *lower row* with vertices $u_0, u_1, \dots, u_{n-1}$ and an *upper row* with vertices $w_0, w_1, \dots, w_{n-1}$. For all $i = 0, 1, \dots, n-2$, there are directed edges $(u_i, u_{i+1})$ and $(w_i, w_{i+1})$, and for all $i = 0, 1, \dots, n-1$, there are edges $\left(u_i, w_{\pi(i)}\right)$ from the lower row to the upper row.

Thus, the only source in $\Delta(n, \pi)$ is $u_0$ and the only sink is $w_{n-1}$. All vertices in the lower row except the leftmost one have indegree 1 and all vertices in the upper row except the leftmost one have indegree 2.

Any DAG of fan-in 2 must have pebbling price at least 3. It is not too hard to see that the graphs $\Delta(n, \pi)$ have pebblings in this minimal space: keeping one pebble on vertex $w_i$ of the upper row, move two pebbles consecutively on the lower row until $u_{\pi^{-1}(i+1)}$ is reached, and then pebble $w_{i+1}$. Generalizing this pebbling strategy leads to the following upper bound on the time-space trade-off for any permutation graph.[5]

---

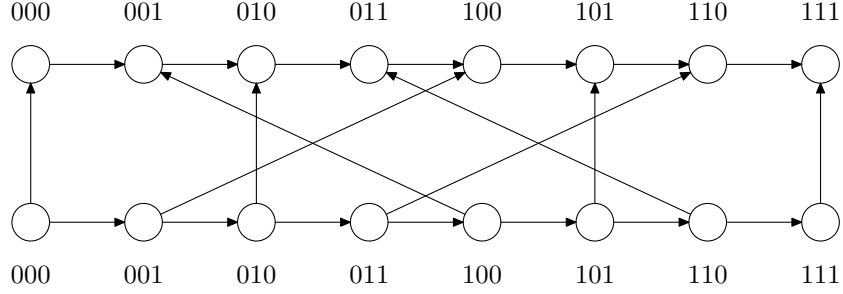

**Figure 6:** Bit reversal graph $\Delta(8, \mathrm{rev})$ on 8 elements.

**Lemma 5.2 ([LT82]).** *Let $\Delta(n, \pi)$ be the permutation graph over $n$ elements for any permutation $\pi$. Then the black pebbling price of $\Delta(n, \pi)$ is $\mathsf{Peb}(\Delta(n, \pi)) = 3$, and for any space $s$, $3 \leq s \leq n$, there is a black pebbling strategy $\mathcal{P}$ for $\Delta(n, \pi)$ with $\mathsf{space}(\mathcal{P}) \leq s$ and $\mathsf{time}(\mathcal{P}) \leq \frac{2n^2}{s-2} + 2n$.*

To prove lower bounds for permutation graphs, Lengauer and Tarjan focus on permutations defined in terms of reversing the binary representation of the integers $\{0, 1, \ldots, n-1\}$ when $n$ is an even power of 2.

**Definition 5.3 (Bit reversal graph ([LT82])).** The *$m$-bit reversal* of $i$, $0 \leq i \leq 2^m - 1$, is the integer $\mathrm{rev}_m(i)$ obtained by writing the $m$-bit binary representation of $i$ in reverse order. The *bit reversal graph* $\Delta(2^m, \mathrm{rev}_m)$ is the permutation graph over $n = 2^m$ with respect to $\mathrm{rev}_m$.

We will denote the bit reversal graph by $\Delta(n, \mathrm{rev})$ for simplicity, implicitly assuming that $n = 2^m$. An example of a bit reversal graph can be found in Figure 6.

For bit reversal graphs, the trade-off in Lemma 5.2 for black pebbling is asymptotically tight.

**Theorem 5.4 ([LT82]).** *Suppose that $\mathcal{P}$ is any complete black pebbling of the bit reversal graph $\Delta(n, \mathrm{rev})$ over $n = 2^m$ elements such that $\mathsf{space}(\mathcal{P}) = s$ for $s \geq 3$. Then $\mathsf{time}(\mathcal{P}) \geq \frac{n^2}{8s}$.*

Note, in particular, that if we want to black-pebble $\Delta(n, \mathrm{rev})$ in linear time, then linear space is needed. The proof of Theorem 5.4 relies on the fact that a black pebbling must always proceed through a graph in topological order. For a black-white pebbling this is no longer true, since pebbles may be placed anywhere at any time. Adjusting the argument used in the proof of Theorem 5.4 accordingly, one instead gets the following, weaker lower bound.

**Theorem 5.5 ([LT82]).** *Let $\mathcal{P}$ be any complete black-white pebbling of $\Delta(n, \mathrm{rev})$ with $\mathsf{space}(\mathcal{P}) = s$ for $s \geq 3$. Then $\mathsf{time}(\mathcal{P}) \geq \frac{n^2}{18s^2} + 2n$.*

When first looking at the proof of Theorem 5.5, it might seem that the bound should not really have to be weaker than in Theorem 5.4 but that this could plausibly be just a consequence of the analysis being harder to carry out in the black-white pebbling case. Somewhat surprisingly, however, Lengauer and Tarjan prove that Theorem 5.5 is in fact tight. That is, one can do (much) better using white pebbles in addition to the black ones. In particular, there is a linear-time black-white pebbling strategy for $\Delta(n, \mathrm{rev})$ using only order of $\sqrt{n}$ pebbles. Moreover, it is possible to transform the pebbling strategy in [LT82] into a bounded labelled pebbling. We conclude our discussion of permutation graphs by stating and proving this as a formal theorem.

**Theorem 5.6.** *Let $\Delta(n, \mathrm{rev})$ be the bit reversal graph over $n = 2^m$ elements. Then for any space parameter $s \geq 3$ there is a complete $(2 + 2s/3, 2)$-bounded labelled pebbling $\mathcal{L}$ of $\Delta(n, \mathrm{rev})$ with $\mathsf{space}(\mathcal{L}) \leq s$ and $\mathsf{time}(\mathcal{L}) \leq 288 \frac{n^2}{s^2} + 22n$.*

Theorem 5.6 is an easy corollary of the next lemma. We establish the lemma first and then explain how it implies the theorem. We also remark that our proof follows [LT82] fairly closely. Thus, our contribution consists in adapting the argument to the bounded labelled pebbling framework.

**Lemma 5.7.** *For all $s$, $3 \le s \le 3\sqrt{n}$, there is a complete $(2 + 2s/3, 2)$-bounded labelled pebbling $\mathcal{L}$ of $\Delta(n, \mathrm{rev})$ with* $\mathrm{space}(\mathcal{L}) \le s$ *and* $\mathrm{time}(\mathcal{L}) \le 288\frac{n^2}{s^2} + 6n$.

*Proof of Lemma 5.7.* Write $m = \log n$ and let $r$ be the non-negative integer such that $3 \cdot 2^r \le s < 3 \cdot 2^{r+1}$. Divide the upper row of $\Delta(n, \mathrm{rev})$ into $2^r$ *intervals*

$$I_j = \{w_{j \cdot 2^{m-r} + k} \mid k = 0, 1, \dots, 2^{m-r} - 1\} \tag{5.1}$$

of size $2^{m-r}$ for $j = 0, \dots, 2^r - 1$ and then subdivide each interval into $2^{m-2r}$ *chunks* by defining

$$C_j^i = \{w_{j \cdot 2^{m-r} + i \cdot 2^r + k} \mid k = 0, 1, \dots, 2^r - 1\} \tag{5.2}$$

for $i = 0, \dots, 2^{m-2r} - 1$. (Note that $2^{m-2r} \ge 1$ since $s \le 3\sqrt{n}$ by assumption.) Figure 7 exemplifies these definitions on the 32-element bit reversal DAG with $2^2$ intervals and 2 chunks per interval.

The pebbling strategy will proceed in $2^{m-2r}$ *phases* corresponding to the $2^{m-2r}$ chunks in each interval, and in $2^r$ *stages* within each phase corresponding to the different intervals. All the phases in the pebbling are completely analogous except for some minor tweaks in the first and final phases, which we refer to as the 0th and $(2^{m-2r} - 1)$st phases, respectively. To help the reader parse the notation, we note that in what follows superscripts $i$ will correspond to phases/chunks and subscripts $j$ to stages/intervals. We reserve $2^r$ independent black pebbles for the lower row, $2^r$ dependent black pebbles for the "current chunks" in the upper row, and $2^r - 1$ supporting white pebbles for theses dependent black pebbles. These white pebbles will be placed on the rightmost vertices in $I_0, I_1, \dots, I_{2^r-2}$. By the way we chose $r$, this leaves one auxiliary pebble to help with advancing the other pebbles.

We start the 0th stage in the 0th phase by doing what is in essence a complete black-only pebbling of the lower row, leaving $2^r$ independent black pebbles on

$$U_0^0 = \{u_{\mathrm{rev}_m(k)}\langle \emptyset \rangle \mid k = 0, 1, \dots, 2^r - 1\} \ . \tag{5.3}$$

More formally, this is done as follows. Introduce the subconfigurations $u_0\langle \emptyset \rangle$ and $u_1\langle u_0 \rangle$, and then merge them to get $u_1\langle \emptyset \rangle$. Next, introduce $u_2\langle u_1 \rangle$ and merge with $u_1\langle \emptyset \rangle$ to get $u_2\langle \emptyset \rangle$. We continue in this way along the lower row, erasing all subconfigurations $u_i\langle u_{i-1} \rangle$ as we go, as well as all subconfigurations $u_i\langle \emptyset \rangle$ not found in $U_0^0$.

Once we have the independent black pebbles in $U_0^0$, we use them to "sweep" a black pebble past the 0th chunk of $I_0$ in the upper row, leaving it on the rightmost vertex $w_{2^r-1}$. In formal notation, we introduce $w_0\langle u_0 \rangle$, merge with $u_0\langle \emptyset \rangle$ to get $w_0\langle \emptyset \rangle$, and then erase $u_0\langle u_0 \rangle$. Next, we introduce $w_1\langle w_0, u_{\mathrm{rev}_m(1)} \rangle$ and merge first with $w_0\langle \emptyset \rangle$ and then with $u_{\mathrm{rev}_m(1)}\langle \emptyset \rangle$, resulting in $w_1\langle \emptyset \rangle$. The dependent black pebbles on $w_1$ are then erased. Next, we introduce $w_2\langle w_1, u_{\mathrm{rev}_m(2)} \rangle$ and merge $w_1\langle \emptyset \rangle$ and $u_{\mathrm{rev}_m(2)}\langle \emptyset \rangle$ to get $w_2\langle \emptyset \rangle$, after which the dependent black pebbles on $w_2$ are erased. Moving right in this fashion, we finally derive $w_{2^r-1}\langle \emptyset \rangle$, noting that all the independent black pebbles $u_{\mathrm{rev}_m(i)}\langle \emptyset \rangle$ that we need for this are present in $U_0^0$. This concludes the 0th stage of our labelled pebbling.

In the next stage, we move all independent black pebbles in $U_0^0$ on the lower row exactly one step to the right to the vertices $u_k$ for $k = 1, \mathrm{rev}_m(1) + 1, \mathrm{rev}_m(2) + 1, \dots, \mathrm{rev}_m(2^r - 1) + 1$. For $k = 1$, this is done by introducing $u_1\langle u_0 \rangle$, merging with $u_0\langle \emptyset \rangle$ to get $u_1\langle \emptyset \rangle$, and then erasing $u_1\langle u_0 \rangle$ and $u_0\langle \emptyset \rangle$. The general case is of course completely analogous. Using the fact that $1 = \mathrm{rev}_m(\mathrm{rev}_r(1) \cdot 2^{m-r})$, we see that we now have independent black pebbles on

$$U_1^0 = \{u_{\mathrm{rev}_m(\mathrm{rev}_r(1) \cdot 2^{m-r} + k)}\langle \emptyset \rangle \mid k = 0, 1, \dots, 2^r - 1\} \ , \tag{5.4}$$

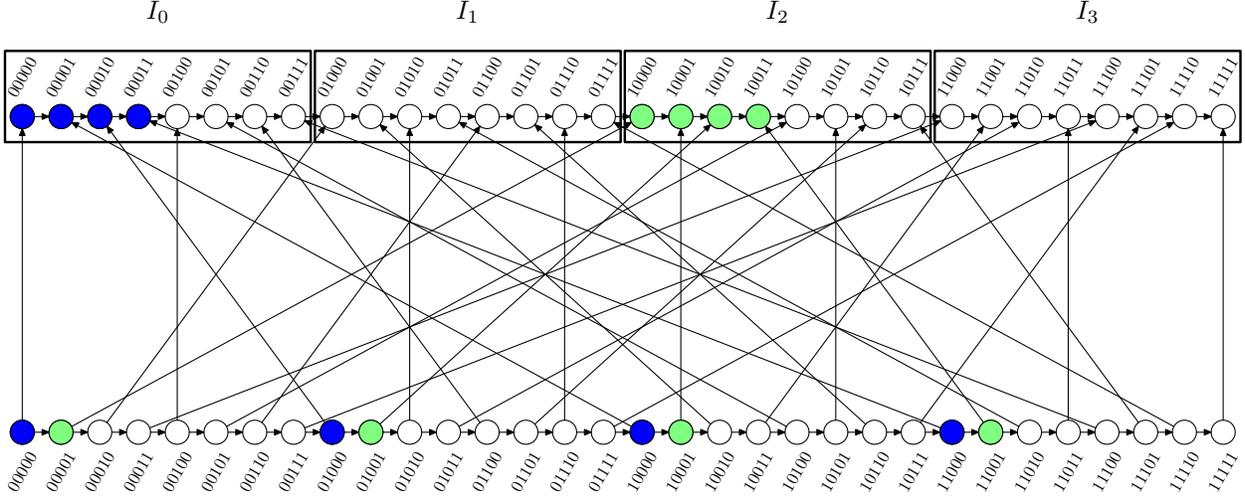

**Figure 7:** Intervals $I_j$ for $r = 2$ in $\Delta(32, \mathrm{rev})$ and 0th chunks in $I_0$ and $I_{\mathrm{rev}_r(1)} = I_2$ with inverse images.

which by (5.2) is the set of all predecessors in the lower row of the 0th chunk $C^0_{\mathrm{rev}_r(1)}$ of the interval $I_{\mathrm{rev}_r(1)}$. This crucial fact is illustrated in Figure 7.

Intuitively, what we want to do now is to place a white pebble on the rightmost vertex of the interval $I_{\mathrm{rev}_r(1)-1}$ and use this white pebble plus the lower-row black pebbles on $U^0_1$ to sweep a black pebble all the way to the rightmost vertex in the 0th chunk of $I_{\mathrm{rev}_r(1)}$. To accomplish this, first introduce $w_{\mathrm{rev}_r(1)\cdot 2^{m-r}}\langle w_{\mathrm{rev}_r(1)\cdot 2^{m-r}-1}, u_{\mathrm{rev}_m(\mathrm{rev}_r(1)\cdot 2^{m-r})}\rangle$ and merge this subconfiguration with the independent black pebble $u_{\mathrm{rev}_m(\mathrm{rev}_r(1)\cdot 2^{m-r})}\langle\emptyset\rangle$, which is present in $U^0_1$, to derive $w_{\mathrm{rev}_r(1)\cdot 2^{m-r}}\langle w_{\mathrm{rev}_r(1)\cdot 2^{m-r}-1}\rangle$. Then introduce $w_{\mathrm{rev}_r(1)\cdot 2^{m-r}+1}\langle w_{\mathrm{rev}_r(1)\cdot 2^{m-r}}, u_{\mathrm{rev}_m(\mathrm{rev}_r(1)\cdot 2^{m-r}+1)}\rangle$ and merge to get the subconfiguration $w_{\mathrm{rev}_r(1)\cdot 2^{m-r}+1}\langle w_{\mathrm{rev}_r(1)\cdot 2^{m-r}-1}\rangle$. Continuing in this way, erasing dependent black pebbles in the upper row as soon as they are no longer needed, we advance a black pebble along all the vertices of the 0th chunk of the interval $I_{\mathrm{rev}_r(1)}$, finally arriving at the pebble subconfiguration $w_{\mathrm{rev}_r(1)\cdot 2^{m-r}+2^r-1}\langle w_{\mathrm{rev}_r(1)\cdot 2^{m-r}-1}\rangle$. This concludes stage 1 of phase 0.

The rest of the stages of phase 0 are completely analogous. In the $j$th stage, we can move the lower-row pebbles from $U^0_{j-1}$ to $U^0_j$ where this notation is generalized to mean

$$U^0_j = \left\{ u_{\mathrm{rev}_m(\mathrm{rev}_r(j)\cdot 2^{m-r}+k)}\langle\emptyset\rangle \;\middle|\; k = 0, 1, \ldots, 2^r - 1 \right\} \tag{5.5}$$

for all $j \leq 2^r - 1$, and then place black pebbles on the rightmost vertex in every chunk $C^0_{\mathrm{rev}_r(j)}$ with the help of a white pebble on the rightmost vertex in $I_{\mathrm{rev}_r(j)-1}$, i.e., , derive pebble subconfigurations $w_{\mathrm{rev}_r(j)\cdot 2^{m-r}+2^r-1}\langle w_{\mathrm{rev}_r(j)\cdot 2^{m-r}-1}\rangle$. At the end of the final stage of phase 0, we thus have black pebbles on the rightmost vertices of all 0th chunks and white pebbles on the rightmost vertices of $I_0, I_1, \ldots, I_{2^r-2}$. Later phases will move the black pebbles to the right, chunk by chunk, while leaving the white pebbles in place. We observe that during phase 0, we made at most $n$ introduction moves and $n$ merger moves on the lower row to get the pebbles into "starting position" $U^0_0$, and then exactly $2^r$ introductions and mergers more on the lower row in each of the other $2^r - 1$ stages.

Inductively, suppose at the beginning of phase $i$ that there are dependent black pebbles on the rightmost vertices in all $(i-1)$st chunks, i.e., subconfigurations $w_{\mathrm{rev}_r(j)\cdot 2^{m-r}+i\cdot 2^r-1}\langle w_{\mathrm{rev}_r(j)\cdot 2^{m-r}-1}\rangle$ for all $j > 0$ and $w_{i\cdot 2^r-1}\langle\emptyset\rangle$ for $j = 0$. Let us extend the lower-row pebble configuration notation above to full generality and define

$$U^i_j = \left\{ u_{\mathrm{rev}_m(\mathrm{rev}_r(j)\cdot 2^{m-r}+i\cdot 2^r+k)}\langle\emptyset\rangle \;\middle|\; k = 0, 1, \ldots, 2^r - 1 \right\} = \left\{ v\langle\emptyset\rangle \,\middle|\, v \in \mathrm{rev}_m^{-1}\!\left(C^i_{\mathrm{rev}_r(j)}\right) \right\}, \tag{5.6}$$

where the second equality is easily verified from (5.2). In stage $0$ of phase $i$, we rearrange the lower-row black pebbles to obtain the configuration in $U_0^i$. Since there are already $2^r$ independent black pebbles present somewhere on the lower row, this can be achieved with at most $n - 2^r$ introductions and mergers (essentially by moving the black pebbles to the closest new position to the right—we refer to [LT82] for the details). This allows us to advance the independent black pebble in $I_0$ on the upper row from the rightmost vertex in chunk $i - 1$ to the rightmost vertex in chunk $i$. Moving the independent black pebbles in $U_0^i$ one step to the right in each following stage to $U_1^i, U_2^i$, et cetera, we can sweep dependent black pebbles across the $i$th chunks of the other intervals $I_j$ in the order $I_{\mathrm{rev}_r(1)}, I_{\mathrm{rev}_r(2)}, \ldots, I_{\mathrm{rev}_r(2^r-1)} = I_{2^r-1}$. All in all, we make at most $(n - 2^r) + (2^r - 1) \cdot 2^r$ introductions and merger moves on the vertices in the lower row during phase $i$ for $i \geq 1$.

In the final $(2^{m-2r} - 1)$st phase, we note that there are supporting white pebbles on the rightmost vertex of the chunk in every interval except $I_{2^r-1}$ (where the rightmost vertex is the sink). Therefore, in every stage except the final one, when we make an introduction move on a rightmost vertex, we merge the introduced subconfiguration with the subconfigurations on its two predecessors of this vertex to remove the white pebble. In the very final stage, we obtain an independent black pebble on $w_{n-1}$. Removing all other pebbles from the DAG, which are all independent black pebbles, we have obtained a complete labelled pebbling of $\Delta(n, \mathrm{rev})$.

The space of this pebbling is $3 \cdot 2^r \leq s$ by construction. All subconfigurations $v\langle W \rangle$ have white support size $|W| \leq 2$, and there are always at most $2 \cdot 2^r \leq 2s/3$ "static" subconfigurations plus 2 auxiliary ones. As to the time bound, it is easy to verify that we make an introduction for each upper row vertex exactly once, and 2 mergers are needed to eliminate the white pebbles in the support of the introduced subconfiguration. The number of introductions and mergers in the lower row is at most $2n + (2^r - 1) \cdot 2^{r+1}$ during phase $0$ and at most $2(n - 2^r) + (2^r - 1) \cdot 2^{r+1}$ for each of the other $2^{m-2r} - 1$ phases, and summing up we get a total of at most

$$
\begin{aligned}
2^{m-2r}\big((2n - 2^{r+1}) + (2^r - 1) \cdot 2^{r+1}\big) + 2^{r+1} + 3n &< 2^{m-2r}\big(2n + 2^{2r+1}\big) + 3n \\
&< \frac{n}{(s/6)^2}\big(2n + 2(s/3)^2\big) + 3n \\
&\leq 144\frac{n^2}{s^2} + 3n
\end{aligned}
\tag{5.7}
$$

introduction and merger moves in total, where we used that $2^{m-2r} \geq 1$, $2^r \leq s/3 < 2^{r+1}$, and $s \leq 3\sqrt{n}$. Multiplying by $2$ to take the removal moves into account gives the time bound stated in the lemma. □

*Proof of Theorem 5.6.* For $s \leq 3\sqrt{n}$ this was proven in Lemma 5.7. To get the statement for $s > 3\sqrt{n}$, use the same pebbling strategy as in the proof of Lemma 5.7 but choose $r$ so that $\sqrt{n}/2 < 2^r \leq \sqrt{n}$. Then the number of chunks $2^{m-2r}$ is at most 2, and the time bound derived from (5.7) reduces to $22n$. □

To obtain the graphs $G_n$ of size $\Theta(n)$ in Lemma 1.3, we set $m = \lceil \log_2 n \rceil$ and let $G_n = \Delta(2^m, \mathrm{rev}_m)$. As noted at the beginning of this section, Theorem 1.4 now follows if we combine Lemma 2.5 with Lemma 5.7.

## 5.2 Bounded Pebblings for Absolute Separations of Pebbling Space

To obtain results for resolution matching the pebbling separations of Lemma 1.5 by [Wil88] and Lemma 1.7 by [KS91], it is sufficient to consider a more general graph family studied in the latter paper. To describe how this graph family is constructed we first need an auxiliary definition.

**Definition 5.8** (*m*-line and $(n, m)$-spiral mesh). An *m-line* is a DAG with vertex set $v_1, v_2, \ldots, v_m$ and edge set $\{(v_i, v_{i+1}) \mid i = 1, 2, \ldots, m - 1\}$.

An $(n, m)$-*spiral mesh* is a DAG on vertices $\{v_{i,j} \mid i \in [n], j \in [m]\}$ with edges $(v_{i,j}, v_{i,j+1})$ for $i \in [n]$ and $j \in [m-1]$, $(v_{i,j}, v_{i+1,j})$ for $i \in [n-1]$ and $j \in [m]$, and $(v_{i,m}, v_{i+1,1})$ for $i \in [n-1]$. The *ith column* of the $(n, m)$-spiral mesh consists of the vertices $v_{i,j}$ for $j \in [m]$.

We now present the three-parameter graph family $\Lambda(p, q, k)$ in [KS91]. The construction is by induction over $q$.

**Definition 5.9** ($\Lambda(p, 0, k)$-**graph**). The graph $\Lambda(p, 0, k)$ is a $(1, p)$-mesh, that is, a $p$-line, the *first row* $f_1, f_2, \ldots, f_p$ and *last row* $l_1, l_2, \ldots, l_p$ of which are both defined to be the vertices of the $p$-line.

For $q > 0$, the graph $\Lambda(p, q, k)$ consists of a number of identical building blocks $N(p, q, k)$, which all contain a copy each of $\Lambda(p, q-1, k)$. In the recursive definitions below, we will be somewhat sloppy with the indices in order not to clutter the notation. In particular, if we wanted to be formally correct, all subgraphs and vertices below should be labelled by their "level of recursion" $q$ within the construction, as well as by a number indicating which of the identical copies on recursion level $q$ the vertex resides in, but we believe that adding these extra indices would lead to more confusion than clarity.

The $N(p, q, k)$-block graph construction, defined next, is illustrated in Figure 8. We remark that this graph has been slightly modified as compared to [KS91].[6]

**Definition 5.10** ($N(p, q, k)$-**block [KS91]**). Suppose that $\Lambda(p, q-1, k)$ has been defined. The *block graph* $N(p, q, k)$, where $k \leq p$, consists of the following components:

- a copy of $\Lambda(p, q-1, k)$ with first row $f_1, f_2, \ldots, f_m$ and last row $l_1, l_2, \ldots, l_m$,

- a $\left((p+1)^2, p\right)$-spiral mesh $B$ on vertices $b_{i,j}$, $i \in \left[(p+1)^2\right]$, $j \in [p]$,

- a $\left((p+1)^3, p\right)$-spiral mesh $A$ on vertices $a_{i,j}$, $i \in \left[(p+1)^3\right]$, $j \in [p]$,

- $k$ copies $R_1, \ldots, R_k$ of a $(p+1)$-line, with the $i$th copy having vertices $r_{i,j}$ for $j \in [p+1]$.

For ease of notation, in what follows we will write $n_b = (p+1)^2$ and $n_a = (p+1)^3$ for the number of rows in $B$ and $A$.

The subgraph components are connected by edges as follows (where we use the notation $(u; v)$ for the edge from $u$ to $v$ for clarity):

- $\left(b_{n_b,j}; f_j\right)$ for $j \in [p]$,

- $\left(b_{n_b,j}; r_{i,p+2-j}\right)$ for $i \in [k]$ and $j \in [p]$,

- $\left(l_j; a_{1,j}\right)$ for $j \in [p]$,

- $\left(l_{\lfloor ip/k \rfloor}; r_{i,1}\right)$ for $i \in [k]$, and

- $\left(r_{i,p+1}; a_{1,j}\right)$ for all $i \in [k]$ and all $j$ such that $(i-1)p/k < j \leq ip/k$.

The $i$th column of $N(p, q, k)$ consists of the $i$th columns of $B$, $\Lambda(p, q-1, k)$, and $A$.

We glue the $N(p, q, k)$-blocks together to form the graph $\Lambda(p, q, k)$ as follows.

---

[6]Again, proofs of the results as stated here can be found in [Nor10a].

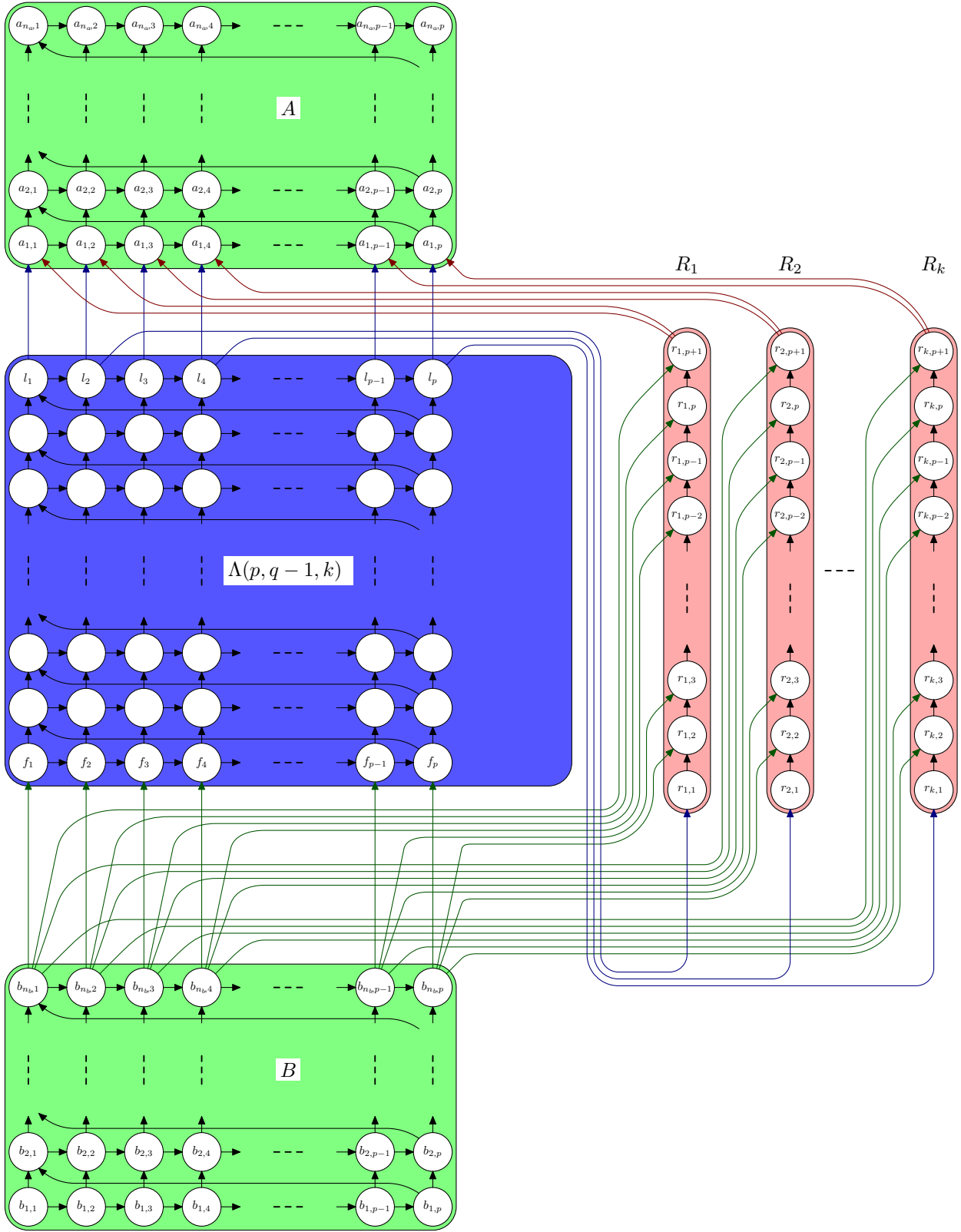

**Figure 8:** Building block $N(p,q,k)$ in graph separating black and black-white pebbling (here $k = p/2$).

**Definition 5.11** ($\Lambda(p,q,k)$-**graph [KS91]**)**.** For $q \leq p$ and $k \leq p$, the graph $\Lambda(p,q,k)$ consists of $\lceil p/k \rceil + 1$ copies of the block graph $N(p,q,k)$, which we denote $N^{(1)}(p,q,k), N^{(2)}(p,q,k), \ldots, N^{(\lceil p/k \rceil + 1)}(p,q,k)$. The edges between the blocks are $\left( a^{(i)}_{n_a,j} ; b^{(i+1)}_{1,j} \right)$ for $i = 1, \ldots, \lceil p/k \rceil$ and $j = 1, \ldots, p$, i.e., the last vertex in every column in the $i$th $N$-block is connected to the first vertex in the same column in the $(i+1)$st $N$-block.

We define the first row $f_1, f_2, \ldots, f_m$ of $\Lambda(p,q,k)$ to consist of the first row $b^{(1)}_{1,1}, b^{(1)}_{1,2}, \ldots, b^{(1)}_{1,p}$, of the first $N$-block and the last row $l_1, l_2, \ldots, l_m$, to consist of the last row $a^{(\lceil p/k \rceil + 1)}_{n_a,1}, a^{(\lceil p/k \rceil + 1)}_{n_a,2}, \ldots, a^{(\lceil p/k \rceil + 1)}_{n_a,p}$ of the last $N$-block. The $i$th column of $\Lambda(p,q,k)$ is defined to be the union of the $i$th columns of all the $N$-blocks.

Let us now first state the properties that we need from the $\Lambda(p,q,k)$-graphs, then show how Lemmas 1.5 and 1.7 and Theorems 1.6 and 1.8 follow from these properties, and finally give the proof that there are efficient bounded labelled pebblings of the graphs.

**Proposition 5.12 ([KS91]).** *The graphs $\Lambda(p,q,k)$, have size $\mathrm{O}\big(\mathrm{poly}(p)(p/k)^q\big)$, maximal vertex indegree 3, and a unique sink.*

**Theorem 5.13 ([KS91]).** *Any complete black pebbling of $\Lambda(p,q,k)$ requires at least $pq$ pebbles.*

**Theorem 5.14.** *Every graph $\Lambda(p,q,k)$ has a complete $(p + kq + 2, 3)$-bounded labelled pebbling.*

If we set $k = p \log \log p / \log p$ and $q = \log p / \log \log p$ in Definition 5.11, it follows from Proposition 5.12 and Theorem 5.13 that we obtain graphs of size polynomial in $p$ with black pebbling price $\Omega(p \log p / \log \log p)$, as claimed in Lemma 1.5. Since these graphs have $(\mathrm{O}(p), \mathrm{O}(1))$-bounded labelled pebblings by Theorem 5.14, we can appeal to Lemma 2.5 to deduce that resolution refutations of pebbling contradictions over these graphs can match the black-white pebbling space bounds, which proves Theorem 1.6. If we instead choose $k = 1$ and $q = p$ in Definition 5.11, we get graphs of size $\exp(\Theta(p \log p))$ that have black pebbling price $\Omega(p^2)$ but admit $(\mathrm{O}(p), \mathrm{O}(1))$-bounded labelled pebblings. This gives us Lemma 1.7 and Theorem 1.8.

Hence, all that remains is to establish Theorem 5.14, and we conclude this section by doing so. Again, we point out that the pebbling strategy presented below follows the one in [KS91] closely, and that our contribution is thus not in designing a completely new pebbling strategy, but in taking an existing strategy and turning it into a bounded labelled pebbling.

Before presenting the formal proof, let us sketch the main idea. Observe that if there were no $R$-graphs in $\Lambda(p,q,k)$ but only the vertices in the $p$ columns, then it would be straightforward to do a complete bottom-up black-only pebbling with just $p + 1$ pebbles. However, this strategy is impossible to implement in the black pebble game. Very briefly, the reason for this is that any black pebbling has to pebble the graph in topological order, but since the predecessors of the vertices in the $R$-graphs have their order reversed—with the source of $R$ having its predecessor in $\Lambda(p, q-1, k)$, whereas the successor vertices have predecessors in the preceding subgraph $B$—this constantly throws the black pebbling off-balance. Using the power of white pebbles, however, we can avoid this problem and place black pebbles on the sinks of all graphs $R_i$, $i \in [k]$, at all levels of recursion in the graph construction, and then do the black bottom-up pebbling of the vertices in the column-part of the graph. The formal details follow.

*Proof of Theorem 5.14.* The labelled pebbling strategy is constructed by induction over $q$. The base case is trivial since $\Lambda(p, 0, k)$ is just a $p$-line. For the the sake of our induction hypothesis, let us do some extra work and note that we can in fact even fill the whole $p$-line with independent black pebbles and still stay within our space bounds. That is, if $l_1, \ldots, l_p$ are the vertices of $\Lambda(p, 0, k)$, we can introduce $l_1 \langle \emptyset \rangle$ and $l_2 \langle l_1 \rangle$ and merge them to get $l_2 \langle \emptyset \rangle$, after which $l_2 \langle l_1 \rangle$ is erased, then introduce $l_3 \langle l_2 \rangle$ and merge with $l_2 \langle \emptyset \rangle$ to obtain $l_3 \langle \emptyset \rangle$, after which $l_3\langle l_2\rangle$ is erased, et cetera, until we have the whole row $\{l_j\langle\emptyset\rangle \mid j \in [p]\}$ of independent black pebbles.

Inductively, suppose that we have constructed for $\Lambda(p, q-1, k)$ a pebbling $\mathcal{L}$ starting with independent black pebbles $\{f_j\langle\emptyset\rangle \mid j \in [p]\}$ on the first row, ending with independent black pebbles $\{l_j\langle\emptyset\rangle \mid j \in [p]\}$ on the last row, and never using more than $p + k(q-1) + 2$ subconfigurations $v\langle W\rangle$ at any time, all with bounded white support size $|W| \le 3$. It is sufficient to construct from $\mathcal{L}$ a labelled pebbling $\mathcal{L}'$ for the block graph $N(p, q, k)$ moving independent black pebbles from the first row of $B$ to the last row of $A$ using no more than $p + kq + 2$ subconfigurations with bounded support size. Such a pebbling is then easily extended to pebbling for all of $\Lambda(p, q, k)$ by pebbling the blocks one by one in a bottom-up fashion. (This is so since we can easily shift independent black pebbles from the last row of an $N$-block to the first row of the next $N$-block using the same kind of labelled pebbling moves that will be discussed more in detail below.)

Thus, suppose that we have independent black pebbles $\{b_{1,j}\langle\emptyset\rangle \mid j \in [p]\}$ on all vertices in the first row of $B$. We move these pebbles up one row as follows. First introduce $b_{2,1}\langle b_{1,1}, b_{1,p}\rangle$ and merge with $b_{1,1}\langle\emptyset\rangle$ and $b_{1,p}\langle\emptyset\rangle$ to get $b_{2,1}\langle\emptyset\rangle$, erasing $b_{1,1}\langle\emptyset\rangle$ and the dependent black pebbles on $b_{2,1}$. Next, introduce $b_{2,2}\langle b_{2,1}, b_{1,2}\rangle$ and merge with $b_{1,2}\langle\emptyset\rangle$ and the newly derived subconfiguration $b_{2,1}\langle\emptyset\rangle$ to get $b_{2,2}\langle\emptyset\rangle$, after which the dependent black pebbles on $b_{2,2}$ are erased, as well as $b_{1,2}\langle\emptyset\rangle$. Continuing in this way, erasing pebble subconfigurations as soon as they are no longer needed and using only 2 auxiliary subconfigurations, we can shift the whole row, and we keep on shifting the pebbles row by row, from left to right for each row, until the last row of $B$ has all vertices covered by independent black pebbles $\{b_{n_b,j}\langle\emptyset\rangle \mid j \in [p]\}$.

Next, we want to place black pebbles on the sinks of all the $R_i$-subgraphs. Fix some $i$ and consider $R_i$. Introduce $r_{i,2}\langle r_{i,1}, b_{n_b,p}\rangle$ and merge with $b_{n_b,p}\langle\emptyset\rangle$ to obtain $r_{i,2}\langle r_{i,1}\rangle$, erasing $r_{i,2}\langle r_{i,1}, b_{n_b,p}\rangle$. Continue by introducing $r_{i,3}\langle r_{i,2}, b_{n_b,p-1}\rangle$ and merging it with $b_{n_b,p-1}\langle\emptyset\rangle$ to obtain $r_{i,3}\langle r_{i,2}\rangle$, and then merge this subconfiguration with $r_{i,2}\langle r_{i,1}\rangle$ to derive $r_{i,3}\langle r_{i,1}\rangle$, where the subconfigurations $r_{i,3}\langle r_{i,2}, b_{n_b,p-1}\rangle$, $r_{i,3}\langle r_{i,2}\rangle$, and $r_{i,2}\langle r_{i,1}\rangle$ are erased as soon as they are no longer needed. Working our way up $R_i$ in this fashion, we finally derive $r_{i,p+1}\langle r_{i,1}\rangle$. Note that we use here that we have all the independent black pebbles $b_{n_b,j}\langle\emptyset\rangle$, $j \in [p]$, available. We repeat these pebbling moves for all the $R_i$-graphs to obtain $\{r_{i,p+1}\langle r_{i,1}\rangle \mid i \in [p+1]\}$. For this part of the pebbling we again use 2 auxiliary subconfigurations, and we end up with a total of $k$ subconfigurations on all the subgraphs $R_i$, $i \in [k]$.

Now, shift the independent black pebbles $\{b_{n_b,j}\langle\emptyset\rangle \mid j \in [p]\}$ from the last row of $B$ to $\{f_j\langle\emptyset\rangle \mid j \in [p]\}$ on the first row of $\Lambda(p, q-1, k)$ (by the same kind of moves that have been described in detail above), and then appeal to the induction hypothesis to obtain a pebbling moving these black pebbles further upward to $\{l_j\langle\emptyset\rangle \mid j \in [p]\}$ on the last row of $\Lambda(p, q-1, k)$. By the induction hypothesis, such a pebbling uses at most $p + k(q-1) + 2$ pebble subconfigurations. We note that adding the $k$ pebble subconfigurations on the $R_i$-subgraphs, the total number of subconfigurations exactly meets the upper bound we are aiming for in the inductive step.

To finish the pebbling of $N(p, q, k)$, we first want to eliminate all the white pebbles on $r_{i,1}$, $i \in [k]$, which is possible since there are (independent) black pebbles on the predecessors of these vertices in the last row of $\Lambda(p, q-1, k)$. Thus, for all $i \in [k]$ in turn, introduce $r_{i,1}\langle l_{\lfloor ip/k\rfloor}\rangle$ and merge $r_{i,p+1}\langle r_{i,1}\rangle$ with the introduced subconfiguration as well as with $l_{\lfloor ip/k\rfloor}\langle\emptyset\rangle$ to derive $r_{i,p+1}\langle\emptyset\rangle$, where we erase $r_{i,1}\langle l_{\lfloor ip/k\rfloor}\rangle$ and $r_{i,p+1}\langle r_{i,1}\rangle$ and any intermediate subconfigurations as soon as they are no longer needed. Next, we shift the black pebbles $\{l_j\langle\emptyset\rangle \mid j \in [p]\}$ from the last row of $\Lambda(p, q-1, k)$ to $\{a_{1,j}\langle\emptyset\rangle \mid j \in [p]\}$ on the first row of $A$. This is done in the same way as previous "shifting" moves, and we use that in addition to the pebbles on the last row of $\Lambda(p, q-1, k)$ we also have independent black pebbles on the sinks of all $R_i$-subgraphs. In this part of the pebbling we will need subconfigurations with white support size 3, since that is the indegree of the vertices in the first row of $A$. When we are done shifting, we erase the pebbles $r_{i,p+1}\langle\emptyset\rangle$ from the sinks of the $R_i$-subgraphs. Finally, we move all the black pebbles in $A$ row by row upward, using 2 auxiliary subconfigurations, until the last row of $A$ has all vertices covered by independent black pebbles. This concludes the inductive step, and the theorem follows. □

# 6  Carlson-Savage Graphs and Strong Dual Trade-offs

In this section, we present a full proof of Theorem 1.9 and show how the Carlson-Savage graphs can be used to obtain graphs with strong dual pebbling trade-offs where the upper bounds are in terms of black pebbling and the lower bounds are in terms of black-white pebbling.

We first list the statements that we want to prove in order to establish Theorem 1.9 in Lemmas 6.1, 6.2, and 6.3 below. Note that the lemmas are stated for the graph family $\Gamma(c, r)$ in Definition 2.6. It is straightforward to translate the lemmas to what is needed for Theorem 1.9 by using the single-sink version of $\Gamma(c, r)$ in Definition 3.8 and appealing to Observation 3.9. Then, we show how these lemmas yield pebbling time-space trade-offs. Finally, we provide the formal proofs of the lemmas.

Let us start by recalling the size and pebbling price bounds.

**Lemma 6.1.** *The graphs $\Gamma(c, r)$ are of size $|V(\Gamma(c, r))| = \Theta(cr^3 + c^3r^2)$, and have black-white pebbling price $\mathsf{BW\text{-}Peb}^\emptyset(\Gamma(c, r)) = r + 2$ and black pebbling price $\mathsf{Peb}^\emptyset(\Gamma(c, r)) = 2r + 1$.*

Note that Lemma 6.1 says that the minimum pebbling space required grows linearly with the recursion depth $r$ but is independent of the number of spines $c$ of the DAG.

Next, we need the fact that there is a linear-time completely black pebbling of $\Gamma(c, r)$ in space linear in $c + r$. This is in fact a strict improvement (though easily obtained) of the corresponding result in [CS82].

**Lemma 6.2.** *The graphs $\Gamma(c, r)$ have persistent black pebbling strategies in simultaneous space $\mathrm{O}(c + r)$ and time linear in the size of the graphs.*

Our main result for the Carlson-Savage graphs is the following trade-off for black-white pebbling, which provides us with a variety of pebbling trade-off results if we choose the parameters $c$ and $r$ appropriately.

**Lemma 6.3.** *Suppose that $\mathcal{P}$ is a complete visiting black-white pebbling of $\Gamma(c, r)$ with*

$$\mathsf{space}(\mathcal{P}) < \mathsf{BW\text{-}Peb}^\emptyset(\Gamma(c, r)) + s = (r + 2) + s$$

*for $0 < s \leq c/8 - 1$. Then the time required to perform $\mathcal{P}$ is lower-bounded by*

$$\mathsf{time}(\mathcal{P}) \geq \left(\frac{c - 2s}{4s + 4}\right)^r \cdot r! \ .$$

Observe that Lemma 6.3 is just a special case of Lemma 2.7, obtained by setting $\mathbb{P}_\sigma = \mathbb{P}_\tau = (\emptyset, \emptyset)$, and we already gave a proof of Lemma 2.7 in Section 2.2, assuming some auxiliary technical lemmas. Hence, for Lemma 2.7 all we need to do is to establish the lemmas stated without proof in Section 2.2.

Before showing any lemmas, however, let us now see how we can prove Theorem 1.10 by appealing to Lemmas 6.1, 6.2, and 6.3.

**Theorem 1.10 (restated).** *Let $g(n)$ be any arbitrarily slowly growing monotone function $\omega(1) = g(n) = \mathrm{O}(n^{1/7})$, and let $\epsilon > 0$ be an arbitrarily small positive constant. Then there is a family of explicitly constructible single-sink DAGs $\{G_n\}_{n=1}^\infty$ of size $\Theta(n)$ with constant vertex indegree such that:*

1. *The graph $G_n$ has black-white pebbling price $\mathsf{BW\text{-}Peb}(G) = g(n) + \mathrm{O}(1)$ and black pebbling price $\mathsf{Peb}(G) = 2 \cdot g(n) + \mathrm{O}(1)$.*

2. *There is a complete black pebbling $\mathcal{P}$ of $G_n$ with $\mathsf{time}(\mathcal{P}) = \mathrm{O}(n)$ and $\mathsf{space}(\mathcal{P}) = \mathrm{O}\left(\sqrt[3]{n/g^2(n)}\right)$*

3. *Any complete black-white pebbling $\mathcal{P}$ of $G_n$ in space at most $\left(n/g^2(n)\right)^{1/3-\epsilon}$ requires pebbling time superpolynomial in $n$.*

*Proof.* Consider the graphs $\Gamma(c, r)$ in Definition 2.6. We want to choose the parameters $c$ and $r$ in a suitable way so that get a family of graphs in size $n = \Theta(cr^3 + c^3 r^2)$ (using the bound on the size of $\Gamma(c, r)$ from Lemma 6.1). If we choose $r = r(n) = g(n)$ for $g(n) = \mathrm{O}(n^{1/7})$, this forces $c = c(n) = \Theta\big(\sqrt[3]{n/g^2(n)}\big)$. Consider the graph family $\{H_n\}_{n=1}^\infty$ defined by $H_n = \Gamma(c(n), r(n))$ as above and let $G_n = \widehat{H_n}$ be the single-sink version of $H_n$. This is a family of single-sink DAGs of size $\Theta(n)$.

By Lemma 6.1 combined with Observation 3.9, it holds that $\mathbf{Peb}(G_n) = g(n) + \mathrm{O}(1)$. Also, the black pebbling of $H_n$ in Lemma 6.2 yields a linear-time pebbling of $G_n$ in space $\mathrm{O}\big(\sqrt[3]{n/g^2(n)}\big)$. Now set the parameter $s$ in Lemma 6.3 to $s = c^{1-\epsilon'}$ for $\epsilon' = 3\epsilon$. Then for large enough $n$ we have $s \le c/8 - 1$ and Lemma 6.3 can be applied. We get that if the pebbling space is less than $\big(n/g^2(n)\big)^{1/3-\epsilon}$, then the required time for the black-white pebbling grows as $\big(\Omega(c^{\epsilon'})\big)^r = \big(\Omega(n/g^2(n))\big)^{\epsilon g(n)}$ which is superpolynomial in $n$ for any $g(n) = \omega(1)$. The theorem follows. $\qquad\square$

We also note that using different parameter settings, we can obtain graphs with very *robust* trade-offs in the sense that the lower bound in the trade-off applies over a very wide space range, namely all the way from $\log n$ up to $\approx \sqrt[3]{n}$.

**Theorem 6.4.** *There is a family of explicitly constructible single-sink DAGs $\{G_n\}_{n=1}^\infty$ of size $\Theta(n)$ with constant vertex indegree such that:*

1. $\mathbf{Peb}(G_n) = \mathrm{O}(\log n)$.

2. *There is a complete black pebbling $\mathcal{P}$ of $G_n$ with $\mathbf{time}(\mathcal{P}) = \mathrm{O}(n)$ and $\mathbf{space}(\mathcal{P}) = \mathrm{O}\left(\sqrt[3]{n/\log^2 n}\right)$.*

3. *There is a constant $K > 0$ such that any complete black-white pebbling $\mathcal{P}$ of $G_n$ in space at most $K\sqrt[3]{n/\log^2 n}$ must take time $n^{\Omega(\log\log n)}$.*

*Proof.* Consider the graphs $\Gamma(c, r)$ in Definition 2.6 with parameters chosen so that $c = 2^r$. Then the size of $\Gamma(c, r)$ is $\Theta(r^2 2^{3r})$ by Lemma 6.1. Let $r(n) = \max\{r : r^2 2^{3r} \le n\}$ and define the graph family $\{G_n\}_{n=1}^\infty$ to be the single-sink version of $\Gamma(2^r, r)$ for $r = r(n)$.

Translating from $G_n$ back to $\Gamma(c, r)$ we have parameters $r = \Theta(\log n)$ and $c = \Theta\big((n/\log^2 n)^{1/3}\big)$, so Lemma 6.1 yields that $\mathbf{Peb}(G_n) = \mathrm{O}(\log n)$. Hence, the linear-time persistent black pebbling of $G_n$ in Lemma 6.2 has space $\mathrm{O}\big((n/\log^2 n)^{1/3}\big)$.

Setting $s = c/8 - 1$ in Lemma 6.3 shows that there is a constant $K$ such that if the space of a black-white pebbling $\mathcal{P}$ drops below $K \cdot (n/\log^2 n)^{1/3} \le (r+2) + s$, then we must have

$$\mathbf{time}(\mathcal{P}) \ge \mathrm{O}(1)^r \cdot r! = n^{\Omega(\log\log n)} \tag{6.1}$$

(where we used that $r = \Theta(\log n)$ for the final equality). The theorem follows. $\qquad\square$

As a final application of Theorem 1.9, we show that it can be used to construct DAGs with not only superpolynomial but even exponential trade-offs. A simple counting argument can be used to show that we can never expect to get exponential trade-offs from DAGs with polylogarithmic pebbling price. However, if we move to graphs with pebbling price $\Omega(n^\epsilon)$ for some constant $\epsilon > 0$, such graphs could potentially exhibit exponential trade-offs. We obtain such a family of graphs by again adjusting the parameters in Definition 2.6 appropriately.

**Theorem 6.5.** *There is a family of explicitly constructible single-sink DAGs $\{G_n\}_{n=1}^\infty$ of size $\Theta(n)$ with constant vertex indegree such that:*

1. $\mathbf{Peb}(G_n) = \mathrm{O}\big(\sqrt[8]{n}\big)$.

2. *There is a complete black pebbling $\mathcal{P}$ of $G_n$ with $\mathsf{time}(\mathcal{P}) = \mathrm{O}(n)$ and $\mathsf{space}(\mathcal{P}) = \mathrm{O}\left(\sqrt[4]{n}\right)$.*

3. *There is a constant $K > 0$ such that any complete black-white pebbling of $G_n$ in space at most $K\sqrt[4]{n}$ must take time $\left(\sqrt[8]{n}\right)!$.*

*Proof.* Use the single-sink version of $\Gamma(c, r)$ as above with parameters $c = \sqrt[4]{n}$ and $r = \sqrt[8]{n}$. $\qquad\blacksquare$

We remark that there is nothing magic in our particular choice of parameters $c$ and $r$ in Theorem 6.5. Other parameters could be plugged in instead and yield slightly different results. Note also that again we have a certain robustness in the trade-off results in that it holds for space from $\sqrt[8]{n}$ to $\sqrt[4]{n}$, at which point it drops sharply to allow a linear-time pebbling.

We now turn to the proofs of Lemmas 6.1, 6.2, and 6.3. In the proofs we will need a few useful auxiliary lemmas, the first of which gives us information about how the spines in the Carlson-Savage DAGs are being pebbled. We will use this information repeatedly in what follows.

**Lemma 6.6 (Rephrasing of Lemma 2.8).** *Suppose that $G$ is a DAG and that $v$ is a vertex in $G$ with a path $Q$ to some sink $z_i \in Z(G)$ such that all vertices in $Q \setminus \{z_i\}$ have outdegree 1. Then any frugal black-white pebbling strategy pebbles $v$ exactly once, and the path $Q$ contains pebbles during one contiguous time interval.*

*Proof.* By induction from the sink backwards. The induction base is immediate. For the inductive step, suppose $v$ has immediate successor $w$ and that $w$ is pebbled exactly once.

If $w$ is black-pebbled at time $\sigma$, then $v$ has been pebbled before this and the first pebble placed on $v$ stays until time $\sigma$. No second placement of a pebble on $v$ after time $\sigma$ can be essential since $v$ has no other immediate successor than $w$. If $w$ is white-pebbled and the pebble is removed at time $\sigma$, then the first pebble placed on $v$ stays until time $\sigma$ and no second placement of a pebble on $v$ after time $\sigma$ can be essential.

Thus each vertex on the path is pebbled exactly once, and the time intervals when a vertex $v$ and its successor $w$ have pebbles on them overlap. The lemma follows. $\qquad\blacksquare$

The second auxiliary lemma speaks about subgraphs $H$ of a DAG $G$ whose only connection to the rest of the graph $G \setminus H$ are via the sink of $H$. Note that the pyramids in $\Gamma(c, r)$ satisfy this condition.

**Lemma 6.7 (Rephrasing of Lemma 2.9).** *Let $G$ be a DAG and $H$ a subgraph in $G$ such that $H$ has a unique sink $z_h$ and the only edges between $V(H)$ and $V(G) \setminus V(H)$ emanate from $z_h$. Suppose that $\mathcal{P}$ is any frugal complete pebbling of $G$ having the property that $H$ is completely empty of pebbles at some given time $\tau'$ but at least one vertex of $H$ has been pebbled during the time interval $[0, \tau']$. Then $\mathcal{P}$ pebbles $H$ completely during the interval $[0, \tau']$.*

*Proof.* Suppose that $v \in V(H)$ is pebbled at time $\sigma' < \tau'$. Note that all paths starting in $v$ must hit $z_h$ sooner or later, since $z_h$ is the unique sink of $H$ and there is no other way out of $H$ into the rest of $G$. Consider the longest path from $v$ to $z_h$. If this path has length 1, clearly $z_h$ must be pebbled before time $\tau'$ since otherwise the pebble placement on $v$ is non-essential. The same statement follows for any $v$ by induction over the path length. But since $H$ is empty at times 0 and $\tau'$ and $z_h$ is pebbled during $(0, \tau')$, $H$ is completely pebbled during this time interval. $\qquad\blacksquare$

Let us now establish that the size and pebbling price of the Carlson-Savage DAGs are as claimed.

*Proof of Lemma 6.1.* The base case graph $\Gamma(c, 1)$ in Definition 2.6 has size $c + 2$. A pyramid of height $h$ has $(h+1)(h+2)/2$ vertices, so the $c$ pyramids of height $2(r-1)$ in $\Gamma(c, r)$ contribute $cr(2r-1)$ vertices.

The $c$ spines with $cr$ sections of $2c$ vertices each contribute a total of $2c^3r$ vertices. And then there are the vertices in $\Gamma(c, r-1)$. Summing up, the total number of vertices in $\Gamma(c, r)$ is

$$(c+2) + \sum_{i=2}^{r} \big(ci(2i-1) + 2c^3 i\big) = \Theta\big(cr^3 + c^3 r^2\big) \tag{6.2}$$

as is stated in the lemma.

Clearly, $\textit{BW-Peb}^{\emptyset}(\Gamma(c,1)) = \textit{Peb}^{\emptyset}(\Gamma(c,1)) = 3$, since pebbling a vertex with fan-in 2 requires 3 pebbles and $\Gamma(c,1)$ can be completely pebbled in this way by placing pebbles on the two sources and then pebbling and unpebbling the sinks one by one.

Suppose inductively that $\textit{BW-Peb}^{\emptyset}(\Gamma(c,r)) = r+2$ and consider $\Gamma(c, r+1)$. It is straightforward to see that $\textit{BW-Peb}^{\emptyset}(\Gamma(c, r+1)) \leq r+3$. Every pyramid $\Pi_{2r}^{(j)}$ can be completely pebbled with $r+2$ pebbles (Theorem 3.15). We can pebble each spine bottom-up in the following way, section by section. Suppose by induction that we have a black pebble on the last vertex $v[i-1]_{2c}$ in the $(i-1)$st section. Keeping the pebble on $v[i-1]_{2c}$, perform a complete visiting pebbling of $\Pi_{2r}^{(1)}$. At some point during this pebbling we must have a pebble on the pyramid sink $z_1$ and at most $r$ other pebbles on the pyramid (by Proposition 3.10). At this time, place a black pebble on $v[i]_1$ and remove the pebble from $v[i-1]_{2c}$. Complete the pebbling of $\Pi_{2r}^{(1)}$, leaving the pyramid empty. Performing complete visiting pebblings of $\Pi_{2r}^{(2)}, \ldots, \Pi_{2r}^{(c)}$ in an analogous fashion allows us to move the black pebble along $v[i]_2, \ldots, v[i]_c$, never exceeding total pebbling space $r+3$. In the same way, for every visiting pebbling $\mathcal{P}$ of $\Gamma(c,r)$ there must exist times $\sigma_i$ for all $i = 1, \ldots, c$, when $\textit{space}(\mathbb{P}_{\sigma_i}) < \textit{space}(\mathcal{P})$ and the sink $\gamma_i$ contains a pebble. Performing a minimum-space pebbling of $\Gamma(c,r)$, possibly $c$ times if necessary, this allows us to advance the black pebble along $v[i]_{c+1}, \ldots, v[i]_{2c}$, never exceeding total pebbling space $r+3$. This shows that $\Gamma(c, r+1)$ can be completely pebbled with $r+3$ pebbles. A simple syntactic adaptation of this arguments for black pebbling (appealing to Theorem 3.15 for the black pebbling price of pyramids) also yields $\textit{Peb}^{\emptyset}(\Gamma(c,r)) \leq 2r+3$.

To prove that there are matching lower bounds for the pebbling constructed above, it is sufficient to show that some pyramid $\Pi_{2r}^{(j)}$ must be completely pebbled while there is at least one pebble on $\Gamma(c, r+1)$ outside of $\Pi_{2r}^{(j)}$. To see why, note that if we can prove this, then simply by using the the fact that $\textit{BW-Peb}^{\emptyset}(\Pi_{2r}) = r+2$ and $\textit{BW-Peb}^{\emptyset}(\Pi_{2r}) = 2r+2$ and adding one for the pebble outside of $\Pi_{2r}^{(j)}$ we have the matching lower bounds that we need. We present the argument for black-white pebbling, which is the harder case. The black-only pebbling case is handled completely analogously.

Suppose in order to get a contradiction that there is a visiting pebbling strategy $\mathcal{P}$ for $\Gamma(c, r+1)$ in space $r+2$. By Observation 3.6, $\mathcal{P}$ performs a complete visiting pebbling of every pyramid $\Pi_{2r}^{(j)}$. Consider the first time $\tau_1$ when some pyramid has been completely pebbled and let this pyramid be $\Pi_{2r}^{(j_1)}$. Then at some time $\sigma_1 < \tau_1$ there are $r+2$ pebbles on $\Pi_{2r}^{(j_1)}$ and the rest of the graph $\Gamma(c, r+1)$ must be empty of pebbles at this point.

We claim that this implies that no vertex in $\Gamma(c, r+1)$ outside of the pyramid $\Pi_{2r}^{(j_1)}$ has been pebbled before time $\sigma_1$. Let us prove this crucial fact by a case analysis.

1. No vertex $v$ in any other pyramid $\Pi_{2r}^{(j')}$ can have been pebbled before time $\sigma_1$. For if so, Lemma 6.7 says that $\Pi_{2r}^{(j')}$ has been completely pebbled before time $\sigma_1$, contradicting our choice of $\Pi_{2r}^{(j_1)}$ as the first such pyramid.

2. No vertex on a spine has been pebbled before time $\sigma_1$. This is so since Lemma 6.6 tells us that if some vertex on a spine has been pebbled, then the whole spine must have been pebbled in view of the fact that it is empty at time $\sigma_1$. But then Lemma 3.12 implies that all pyramid sinks must have been pebbled. This case has already been excluded.

3. Finally, no vertex $v$ in $\Gamma(c, r)$ can have been pebbled before time $\sigma_1$. Otherwise the frugality of $\mathcal{P}$ implies (by pattern matching on the arguments in the proofs of Lemmas 3.12 and 6.6) that some successor of $v$ must have been pebbled as well, and some successor of that successor et cetera, all the way up to where $\Gamma(c, r)$ connects with the spines. But we have ruled out the possibility that a spine vertex has been pebbled.

This establishes the claim, and we are now almost done. To clinch the argument, we need a couple of final observations. Note first that by frugality, at some time in the interval $(\sigma_1, \tau_1)$ some vertex in some spine must have been pebbled. This is so since the pyramid sink $z_{j_1}$ has been pebbled before time $\tau_1$, all of $\Pi_{2r}^{(j_1)}$ is empty at time $\tau_1$, and all spines are empty at time $\sigma_1 < \tau_1$. But then Lemma 6.6 tells us that there will remain a pebble on this spine until all of the spine has been completely pebbled.

Consider now the second pyramid $\Pi_{2r}^{(j_2)}$ completely pebbled by $\mathcal{P}$, say, at time $\tau_2$. At some point in time $\sigma_2 < \tau_2$ we have $r + 2$ pebbles on $\Pi_{2r}^{(j_2)}$, and moreover $\sigma_2 > \tau_1$ since $\Pi_{2r}^{(j_2)}$ is empty at time $\tau_1$. But now it must hold that either there is a pebble on a spine at this point, or, if all spines are completely empty, that some spine has been completely pebbled. If some spine has been completely pebbled, however, this in turn implies (appealing to Lemma 3.12 again) that there must be some pebble somewhere in some other pyramid $\Pi_{2r}^{(j')}$ at time $\sigma_2$. Thus the pebbling space exceeds $r + 2$ and we have obtained our contradiction. The lemma follows. $\qquad\blacksquare$

Studying the pebbling strategies in the proof of Lemma 6.1, it is not hard to see that they are very inefficient. The subgraphs in $\Gamma(c, r)$ will be pebbled over and over again, and for every step in the recursion the time required multiplies. We next show that if we are a bit more generous with the pebbling space, then we can get down to linear time.

*Proof of Lemma 6.2.* We want to prove that $\Gamma(c, r)$ has a persistent black pebbling strategy $\mathcal{P}$ that pebbles every vertex in $\Gamma(c, r)$ exactly once and uses space $\mathrm{O}(c + r)$. Suppose that there is such a pebbling strategy $\mathcal{P}_r$ for $\Gamma(c, r)$. We describe how to construct a pebbling $\mathcal{P}_{r+1}$ for $\Gamma(c, r + 1)$ inductively. Note that the base case for $\Gamma(c, 1)$ is trivial.

The construction of $\mathcal{P}_{r+1}$ is very straightforward. First use $\mathcal{P}_r$ to make a persistent pebbling of $\Gamma(c, r)$ in space $\mathrm{O}(c + r)$. At the end of $\mathcal{P}_r$, we have $c$ pebbles on the sinks $\gamma_1, \ldots, \gamma_c$. Keeping these pebbles in place, pebble the pyramids $\Pi_{2r}^{(1)}, \ldots, \Pi_{2r}^{(c)}$ persistently one by one in space $\mathrm{O}(r)$ with a strategy pebbling each vertex exactly once (for instance, by pebbling the pyramid bottom-up level by level). We leave pebbles on all pyramid sinks $z_1, \ldots, z_c$. Thus the pebbling only requires space $\mathrm{O}(c + r)$ and at the end we have $2c$ black pebbles on all pyramid sinks $z_1, \ldots, z_c$ and all sinks $\gamma_1, \ldots, \gamma_c$ of $\Gamma(c, r)$. Keeping all these pebbles in place, we can pebble all $c$ spines in parallel in linear time with $c + 1$ extra pebbles. $\qquad\blacksquare$

It remains to fill in the gaps in the proof of Lemma 2.7 and its special case Lemma 6.3. Recall that the proof of Lemma 2.7 presented in Section 2.2 hinged on the claims that not too many pyramids can be pebbled simultaneously in a space-efficient pebbling, and that this is true for the spines as well. Assuming these two claims, we could show that that as any spine was pebbled, the pebbling had to alternate back and forth between time intervals when there are a lot of pebbles on some pyramid and time intervals when all sinks in $\Gamma(c, r)$ are pebbled. This allowed us to apply the induction hypothesis multiple times and obtain the required lower bound.

Hence, all that remains to complete the proof of Lemma 2.7 is to establish the two technical lemmas that upper-bound how many pyramids and spine sections can contain pebbles simultaneously at any one given time in a pebbling subjected to space constraints as in Lemma 2.7. The claims in the two lemmas are very similar in spirit, as are the proofs, so we state the lemmas together and then present the proofs in sequence.

**Lemma 6.8 (Rephrasing of Lemma 2.10).** *Suppose that $\mathcal{P} = \{\mathbb{P}_\sigma, \ldots, \mathbb{P}_\tau\}$ is a conditional black-white pebbling on $\Gamma(c, r)$ and that $s$ is a constant satisfying the conditions in Lemma 2.7. Then at all times during the pebbling $\mathcal{P}$ strictly less than $4(s + 1)$ pyramids $\Pi_{2r}^{(j)}$ contain pebbles simultaneously.*

**Lemma 6.9 (Rephrasing of Lemma 2.11).** *Suppose that $\mathcal{P} = \{\mathbb{P}_\sigma, \ldots, \mathbb{P}_\tau\}$ is a conditional black-white pebbling on $\Gamma(c, r)$ and that $s$ is a constant satisfying the conditions in Lemma 2.7. Then at all times during the pebbling $\mathcal{P}$ strictly less than $4(s + 1)$ spine sections contain pebbles simultaneously.*

Note that Lemma 6.9 provides a total bound on the number of pebbled sections in all $c$ spines. There might be spines containing several sections being pebbled simultaneously (in fact, this is exactly what one would expect a black-white pebbling to do to optimize the time given the space constraints), but what Lemma 6.9 says is that if we fix an arbitrary time $t \in [\sigma, \tau]$, add up the number of sections containing pebbles at time $t$ in each spine, and sum over all spines, we never exceed $4(s + 1)$ sections in total.

*Proof of Lemma 6.8.* Suppose that on the contrary, there is some time $t^* \in (\sigma, \tau)$ when at least $4s + 4$ pyramids $\Pi^{(j)}$ in $\Gamma(c, r)$ contain pebbles. Of these pyramids, at least $2s + 4$ are empty both at time $\sigma$ and at time $\tau$ since $\textbf{\textit{space}}(\mathbb{P}_\sigma) < s$ and $\textbf{\textit{space}}(\mathbb{P}_\tau) < s$. By Lemma 6.7, these pyramids, which we denote $\Pi^{(1)}, \ldots, \Pi^{(2s+4)}$, are completely pebbled during $[\sigma, \tau]$. Moreover, we can conclude that for every $\Pi^{(j)}$, $j = 1, \ldots, 2s + 4$, there is an interval $[\sigma_j, \tau_j] \subseteq [\sigma, \tau]$ such that $t^* \in (\sigma_j, \tau_j)$ and $\Pi^{(j)}$ is empty at times $\sigma_j$ and $\tau_j$ but contains pebbles throughout the interval $(\sigma_j, \tau_j)$ during which it is completely pebbled.

For each $\Pi^{(j)}$ there must exist some time $t_j^* \in (\sigma_i, \tau_i)$ when there are at least $r + 1 = \textbf{\textit{BW-Peb}}^\emptyset(\Pi^{(j)})$ pebbles. Fix such a time $t_j^*$ for every pyramid $\Pi^{(j)}$ and assume that all $t_j^*$, $j = 1, \ldots, 2s + 4$, are sorted in increasing order. We have two possible cases:

1. At least half of all $t_j^*$ occur before (or at) time $t^*$, i.e., they satisfy $t_j^* \leq t^*$. If so, look at the largest $t_j^* \leq t^*$. At this time there are at least $r + 1$ pebbles on $\Pi^{(j)}$ and at least $\frac{2s+4}{2} - 1 = s + 1$ pebbles on other pyramids, which means that $\textbf{\textit{space}}(\mathbb{P}_{t_j^*}) \geq (r + 2) + s$. In other words, $\mathcal{P}$ exceeds the space restrictions in Lemma 2.7. Contradiction.

2. At least half of all $t_j^*$ occur after time $t^*$, i.e., they satisfy $t_j^* > t^*$. If we consider the smallest $t_j^*$ larger than $t^*$ we can again conclude that $\textbf{\textit{space}}(\mathbb{P}_{t_j^*}) \geq (r + 1) + (s + 1)$, which is again a contradiction.

Hence, if $\mathcal{P}$ is a pebbling that complies with the restrictions in Lemma 2.7, it can never be the case that $4s + 4$ pyramids $\Pi^{(j)}$ in $\Gamma(c, r)$ contain pebbles simultaneously. $\qquad\blacksquare$

*Proof of Lemma 6.9.* Suppose that at some time $t^* \in (\sigma, \tau)$ at least $4s + 4$ sections contain pebbles. At least $2s + 4$ of these sections are empty at times $\sigma$ and $\tau$. Let us denote these sections $R_1, \ldots, R_{2s+4}$. Appealing to Lemma 6.6, we conclude that there are intervals $[\sigma_j, \tau_j] \subseteq [\sigma, \tau]$ for $j = 1, \ldots, 2s + 4$, such that $t^* \in (\sigma_j, \tau_j)$ and $R_j$ is empty at times $\sigma_j$ and $\tau_j$ but contains pebbles throughout the interval $(\sigma_j, \tau_j)$ during which it is completely pebbled.

By Lemma 6.8, we know that less than $4s + 4$ pyramids contain pebbles at time $\sigma_j$ and similarly at time $\tau_j$. Since all $c$ pyramids in $\Gamma(c, r)$ must have their sinks pebbled during $(\sigma_j, \tau_j)$ but it holds that $2 \cdot (4s + 4) < c$ by the assumptions in Lemma 2.7, we conclude from Lemma 6.7 that for every section $R_j$ we can find some pyramid $\Pi^{(j)}$ that is completely pebbled during the interval $(\sigma_j, \tau_j)$. This, in turn, implies that there is some time $t_j^* \in (\sigma_j, \tau_j)$ when the pyramid $\Pi^{(j)}$ contains at least $\textbf{\textit{BW-Peb}}^\emptyset(\Pi^{(j)}) = r + 1$ pebbles. (We note that many $t_j^*$ can be equal and even refer to the same pyramid, but this is not a problem.)

As in the proof of Lemma 6.8, we now sort the $t_j^*$, $j = 1, \ldots, 2s + 4$, in increasing order and consider the two possible cases. If at least half of all $t_j^*$ satisfy $t_j^* \leq t^*$, we look at the largest $t_j^* \leq t^*$. At this time there are at least $r + 1$ pebbles on $\Pi^{(j)}$ and at least $\frac{2s+4}{2} = s + 2$ pebbles on different sections, which means that $space\left(\mathbb{P}_{t_j^*}\right) \geq r + s + 3$ exceeds the space restrictions. If, on the other hand, at least half of all $t_j^*$ satisfy $t_j^* > t^*$, then for the smallest $t_j^*$ larger than $t^*$ we can again conclude that $space\left(\mathbb{P}_{t_j^*}\right) \geq r + s + 3$, which is a contradiction. The lemma follows. □

As we discussed at the start of this section, Theorem 1.9 now follows by applying Observation 3.9 on the single-sink version of $\Gamma(c, r)$.

As a final note, we remark that not only do our proofs get much more involved when going from the black-only pebbling trade-off in [CS82] to our black-white pebbling trade-off, but the added complications also lead to our bound for black-white pebbling being slightly worse than the one in [CS82] for black pebbling. More specifically, Carlson and Savage are able to prove their results for DAGs having only $\Theta(r)$ sections per spine, whereas we need $\Theta(cr)$ sections in $\Gamma(c, r)$. This blows up the number of vertices, which in turn weakens the trade-offs measured in terms of graph size. It would be interesting to find out whether our proof could in fact be made to work for graphs with only $O(r)$ sections per spine. If so, this would immediately improve the trade-offs for the graphs in Theorems 1.10, 6.4, and 6.5, as well as the resolution trade-offs derived from these graphs in [BN09b].

# 7 Concluding Remarks

It is known that the black-white pebbling price is always a lower bound on the resolution space of refuting pebbling contradictions $Peb_G[f]$ with respect to the "right" functions $f$, as proven in [BN08]. Also, for all graphs studied in this context so far there have been shown to exist refutations of the corresponding pebbling contradictions in space upper-bounded by the black-white pebbling price—trivially for graphs where the black and black-white pebbling prices coincide, and more interestingly for the graphs in the current paper where the black-white pebbling price is asymptotically smaller than the black pebbling price. This naturally raises the question whether it holds in general that the refutation space of pebbling contradictions is asymptotically equal to the black-white pebbling price of the underlying graphs.

**Open Question 1.** *Is in true for any DAG $G$ with bounded vertex indegree and any (fixed) Boolean function $f$ that the pebbling contradiction $Peb_G[f]$ can be refuted in total space $O(\textbf{BW-Peb}(G))$?*

More specifically, one could ask—as a natural first line of attack if one wants to investigate whether the answer to the above question could be yes—if it holds that bounded labelled pebblings are in fact as powerful as general black-white pebblings. In a sense, this is asking whether only a very limited form of nondeterminism is sufficient to realize the full potential of black-white pebbling.

**Open Question 2.** *Does it hold that any complete black-white pebbling $\mathcal{P}$ of a single-sink DAG $G$ with bounded vertex indegree can be simulated by a $(O(space(\mathcal{P})), O(1))$-bounded pebbling $\mathcal{L}$?*

Note that a positive answer to this second question would immediately imply a positive answer to the first question as well by Lemma 2.5.

We have no strong intuition either way regarding Open Question 1, but as to Open Question 2 it would perhaps be somewhat surprising if bounded labelled pebblings turned out to be as strong as general black-white pebblings. Interestingly, although the optimal black-white pebblings of the graphs in Lemma 1.7 can be simulated by bounded pebblings, the same approach does *not* work for the original graphs separating black-white from black-only pebbling in [Wil88]. Indeed, these latter graphs might be a candidate graph family for answering Open Question 2 in the negative, i.e., showing that standard black-white pebblings can be asymptotically stronger than bounded labelled pebblings.

Finally, we are intrigued by the question of whether the properties of the formulas $Peb_G[f]$ shown to hold in [BN08, BN09b] for "the right kind" of functions $f$ in fact extend to the simpler formulas $Peb_G[\lor]$ defined in terms of non-exclusive or.

**Open Question 3.** *Is it true for any DAG G that any resolution refutation $\pi$ of $Peb_G[\vee]$ can be translated into a black-white pebbling of G with time and space upper-bounded asymptotically by the length and space of $\pi$?*

Earlier results in [Nor09, NH08b] can be interpreted as indicating that this should be the case, but the results there only apply to limited classes of graphs and only capture space lower bounds, not time-space trade-offs. And the papers [BN08, BN09b] do not shed any light on this question, as the techniques used there inherently cannot work for formulas defined in terms of non-exclusive or.

If the answer to Open Question 3 is yes—which we would cautiously expect it to be—then this could be useful for settling the complexity of decision problems for resolution proof space, i.e., the problem given a CNF formula $F$ and a space bound $s$ to determine whether $F$ has a resolution refutation in space at most $s$. Reducing from pebbling space by way of formulas $Peb_G[\vee]$ would avoid the blow-up of the gap between upper and lower bounds on pebbling space that cause problems when using, for instance, exclusive or.

## Acknowledgements

I would like to thank David Carlson, Nicholas Pippenger, and John Savage for helpful correspondence regarding their papers on pebbling, and Eli Ben-Sasson and Johan Håstad for stimulating discussions about pebbling and proof complexity. I am also grateful to the *CCC '10* anonymous referees for useful suggestions and comments.